\documentclass [a4paper,oneside] {article}

\usepackage[ansinew]{inputenc}

\usepackage[french]{babel}

\usepackage{amssymb,amsmath, amsfonts}

\usepackage{graphicx,color}

\usepackage{amsthm}

\usepackage{hyperref}

\usepackage{MnSymbol}

\newtheorem{prop} {Proposition}

\theoremstyle{definition}
\newtheorem{df}{Définition} 
\newtheorem*{df*}{Définition}

\theoremstyle{remark}
\newtheorem{rmq}{Remarque}
\newtheorem{example}{Exemple} 
\newtheorem{exm}[example]{Exemple}
\newtheorem{xrc}{Exercice}

\author{Stéphane \textsc{Dugowson}
\footnote{stephane.dugowson@supmeca.com}
}

\title {Structures connectives de l'intrication quantique}

\begin{document}

\maketitle

\begin{abstract}

Dans ce texte, après des rappels d'une part sur la notion de structure connective et d'autre part sur les formalismes de la mécanique quantique, nous associons certaines familles de structures connectives à tout état quantique intriquant un nombre fini quelconque de particules, ainsi qu'à tout \og dispositif de mesure\fg\, portant sur de tels états. Cela nous permet finalement de définir un nouvel outil de classification de l'intrication quantique : l'ordre connectif. 

\mbox{}

\noindent \textit{Mots clés}: Connectivité. Intrication quantique. Borroméen. Ordre connectif.\\

\textbf{Abstract}. \emph{Connectivity structures of quantum entanglement}. In this paper, after some recalls about connectivity structures and about the formalisms of quantum mechanics, we associate some families of connectivity structures with any entangled quantum state, and with any ``measurement device" on such states. This finally allows us to define a new classification tool for quantum entanglement: the connectivity order.

\mbox{}

\noindent \textit{Keywords}: Connectivity. Quantum entanglement. Borromean. Connectivity order.\\

\noindent Mathematics Subject Classification 2010: 54A05, 57M25, 81P40.

\end{abstract}

%
%

\section*{Introduction}


Le présent article s'inscrit dans le cadre général d'une recherche de l'auteur concernant les structures connectives des interactions entre plusieurs entités, et concerne plus spécifiquement les structures connectives susceptibles de décrire l'intrication quantique. Il accompagne et développe les considérations présentées par l'auteur dans plusieurs exposés : le 18 juin 2017 au séminaire de logique catégorique dirigé par Anatole Khélif à l'université Paris Diderot (\cite{Dugowson:20140618}), le 25 juin 2014 à Aix-en-Provence au cours du \emph{Workshop on Diffeology, etc.} organisé par Patrick Iglesias-Zemmour\footnote{\url{http://math.huji.ac.il/~piz/Site/Worshop\%20Diffeology\%202014.html}} ainsi que le 2 juillet 2014 au séminaire du laboratoire Quartz organisé à Supméca (Paris).

\`{A} l'origine de notre intérêt actuel pour le thème de l'intrication quantique, il y a notamment des discussions avec Anatole Khélif et Christophe Chalons dans le cadre du séminaire de logique catégorique de l'Université Paris Diderot, discussions portant en particulier sur les états intriqués et les expériences dites $GHZ$ en référence aux travaux initiés à partir de 1989 par Greenberger, Horne et Zeilinger \cite{GHZ:1989}, états et expériences qui, en mettant en jeu trois, quatre ou davantage de particules, renouvellent le thème de l'intrication quantique initié par Einstein, Podolsky et Rosen ($EPR$). Du point de vue connectif, ce type d'intrication soulève immédiatement la question de savoir dans quelle mesure elle pourrait ou non être dite \emph{borroméenne}\footnote{La structure borroméenne est l'une des structures connectives les plus fondamentales, voir plus loin la section \ref{section rappels sur les structures connectives}.}, et plus généralement de définir la structure connective de tout état quantique intriqué. Se posera alors la question de la réalisabilité quantique de \emph{toute} structure connective finie.

En général, l'intuition initiale de ce que devrait être la structure connective d'une interaction semble \emph{a priori} assez claire : si un sous-ensemble de l'ensemble de toutes les entités parmi celles que l'on considère peut être séparé par la pensée en deux parties entre lesquelles il n'y a pas d'interactions, alors ce sous-ensemble d'entités n'est pas connecté. La structure connective est alors celle qu'engendrent les sous-ensembles connectés\footnote{Bien entendu, on peut imaginer que la structure connective d'une interaction évolue au cours du temps en fonction des interactions elles-mêmes, ou des interactions avec d'autres systèmes.}. Une des difficultés, en particulier dans le cas de l'intrication quantique, est toutefois de savoir quel genre de système constitue un tel sous-ensemble. Comme nous le verrons, cette difficulté provient, au moins dans le cadre du formalisme quantique en termes de vecteurs d'état (par oppositions aux opérateurs de densité), du fait qu'un sous-système quantique \emph{n'est pas} un système quantique (fermé).

Dans le cas d'un entrelacs, plutôt que de considérer d'emblée un sous-entrelacs en \emph{oubliant} les autres composantes, on peut voir un tel sous-entrelacs comme le résultat d'opérations de \emph{coupures} sur les composantes qui ne lui appartiennent pas. Or, de telles coupures ont un analogue dans le formalisme quantique des vecteurs d'état, à savoir la transformation des états consécutive à une expérience de mesure quantique, notamment une mesure projective.

En 1997, Aravind \cite{Aravind:1997}, adoptant ce point de vue, a remarqué que l'état $GHZ$ à trois particules pouvait effectivement être associé à l'entrelacs borroméen... entre autres entrelacs. En 2007, Sugita \cite{Sugita:2007}, s'appuyant cette fois sur le formalisme des opérateurs de densité, a montré que la nature borroméenne de l'état $GHZ$ pouvait effectivement être considérée comme plus fondamentale que les autres descriptions par entrelacs. 
Aucun de ces deux auteurs, toutefois, ne disposait du point de vue connectif --- certes lié aux entrelacs, notamment grâce au théorème de Brunn-Debrunner-Kanenobu (voir \cite{Dugowson:201012} et \cite{Dugowson:201203}) --- et n'a considéré en toute généralité l'intrication quantique d'un nombre quelconque de particules.   

Après quelques rappels sur les structures connectives (section \ref{section rappels sur les structures connectives}), suivis d'un récapitulatif sur les formalismes de la mécanique quantique et en particulier de l'intrication quantique (section \ref{section formalismes quantiques}), c'est donc ce que nous nous proposons de faire dans la section suivante, intitulée \emph{Structures connectives des états quantiques intriqués}. 

La section \ref{section structure relationnelle des dispositifs}, intitulée \emph{Structures relationnelles des dispositifs multilocaux}, vise également à définir des structures connectives, non plus pour les états quantiques intriqués, mais cette fois pour les \emph{expériences de mesure} que l'on peut réaliser sur de tels états, ces expériences étant identifiées à des familles de questions qui peuvent être posées, localement, aux systèmes quantiques considérés.
Afin de pouvoir préciser les structures connectives en question, nous commencerons par formaliser ce type de situations grâce à la notion de \textit{dispositif multilocal} --- notion directement empruntée, avec quelques changements terminologiques, aux relations considérées par Christophe Chalons dans son travail sur les degrés ludiques (voir \cite{Chalons:201405}) ---  puis nous préciserons diverses notions de localité susceptibles de s'appliquer ou non à ces dispositifs. 
  Plusieurs exemples de tels dispositifs et de leurs structures connectives sont alors présentés, portant notamment sur les états quantiques $EPR$ et $GHZ$. 
  

Enfin, l'aspect probabiliste des dispositifs de mesure ayant été laissé de coté dans les considérations précédentes, une dernière section présente rapidement, à titre de piste de recherche, la notion de structure connective d'une famille de variables aléatoires, due à Anatole Khélif.

On trouvera à la fin du texte un indexe des notations, la liste des références bibliographiques et une table des matières.

\section{Rappels sur les structures connectives}\label{section rappels sur les structures connectives}

Nous donnons ci-après quelques rapides rappels sur les structures connectives, le lecteur souhaitant approfondir la question étant invité à se reporter aux références \cite{Dugowson:201012,Dugowson:201203, Dugowson:201306}.

\subsection{Structures connectives intègres}

\begin{df} [Espaces connectifs] Un \emph{espace connectif} est un  couple $X=(\vert X\vert,\kappa(X))$ formé d'un ensemble $\vert X\vert$ et d'un ensemble non vide $\kappa(X)$ de parties de $\vert X\vert$ tel que pour toute famille $\mathcal{I}\in \mathcal{P}(\kappa(X))$, on ait
\begin{displaymath}
\bigcap_{K\in\mathcal{I}}K\ne\emptyset\Longrightarrow \bigcup_{K\in\mathcal{I}}K\in\kappa(X).
\end{displaymath}
\noindent L'ensemble $ \vert X \vert$ est le \emph{support} de $X$, l'ensemble $\kappa(X)$ est la \emph{structure connective} de $X$ et ses élé\-ments sont les \emph{parties connexes} ou \emph{parties connectées} de l'espace connectif $X$. 
Un point $x\in\vert X\vert$ est \emph{absent} s'il n'appartient à aucune partie connexe de $X$, il est \emph{présent} dans le cas contraire. On appelle \emph{composante absente} de $X$ l'ensemble des points absents de $X$. On appelle \emph{composantes connexes} de $X$ les parties connexes maximales pour l'inclusion. Pour tout point présent $x$ de $X$, on appelle \emph{composante connexe de $x$} l'unique composante connexe de $X$ contenant $x$.
Nous dirons d'un espace connectif qu'il est \emph{intègre} si tout  singleton est connecté.
Un \emph{morphisme connectif}, ou \emph{application connective}, d'un espace connectif $(\vert X\vert ,\kappa(X))$ vers un autre $(\vert Y\vert,\kappa(Y))$ est une application $f:\vert X\vert \to \vert Y\vert $ telle que :
\begin{displaymath}
\forall K\in \kappa(X), f(K)\in \kappa(Y).
\end{displaymath}
\end{df}

Dans le présent article, nous ne considérerons que des espaces connectifs finis intègres.

\subsection{Exemples}

\begin{exm}[Espaces connectifs topologiques]  On définit un foncteur $U_T:\mathbf{Top}\to\mathbf{Cnct}$ en associant à tout espace topologique l'espace connectif intègre ayant les mêmes points, et dont les connexes sont les parties connexes pour la topologie de l'espace considéré. Nous dirons qu'un espace connectif est topologique s'il est dans l'image objet de $U_T$.
\end{exm}
 
\begin{exm}[Espaces connectifs graphiques]
La notion de connexité d'un ensemble de sommets d'un graphe  simple  non orienté conduit à la définition d'un foncteur $U_G:\mathbf{Grf}\rightarrow \mathbf{Cnct}$ défini sur la catégorie $\mathbf{Grf}$ dont les objets sont les graphes simples non orientés et dont les flèches sont les applications qui préservent cette connexité. A tout graphe de ce genre, le foncteur $U_G$ associe l'espace connectif ayant pour points les sommets du graphe, et pour parties connexes tout ensemble de sommets connexe au sens des graphes, \textit{i.e.} tel qu'il existe un chemin dans cet ensemble formé d'arêtes reliant de proche en proche toute paire de points de cet ensemble.
Nous dirons qu'un espace connectif est graphique s'il est dans l'image objet de $U_G$.
\end{exm}

\begin{exm}[Espace borroméen, espaces brunniens]\label{exm espace borromeen}
L'exemple le plus simple d'espace intègre ni topologique ni graphique est l'\textit{espace borroméen}  $\mathcal{B}_3$, où pour tout entier $n$ on désigne par $\mathcal{B}_n$  l'espace intègre à $n$ points dont la seule partie connexe non réduite au vide ou à un singleton est la partie pleine :
\[ \vert\mathcal{B}_n\vert=\{0, 1, ..., n-1\} 
\quad\mathrm{ et }\quad \kappa^\bullet(\mathcal{B}_n) =\{\vert\mathcal{B}_n\vert\}, \]
\noindent où, pour tout espace connectif $X$, on a posé \[\kappa^\bullet(X)=\{K\in\kappa(X), card(K)\geqslant 2\}.\]
\end{exm}

\begin{exm}[Structure connective des entrelacs]\label{exm structure connective entrelacs}
\`{A} tout entrelacs $L$ dans $\mathbf{R}^3$ ou $\mathbf{S}^3$, on associe un espace connectif intègre $S_L$ en prenant pour points de $S_L$ les composantes de $L$, les parties connexes de l'espace étant données par les sous-entrelacs de $L$ non séparables par un plan topologique. 
\end{exm}

\subsection{Théorème de Brunn-Kanenobu et conjectures brunniennes}\label{subsection conjectures brunniennes}

L'exemple \ref{exm structure connective entrelacs} admet une sorte de réciproque, énoncée par Hermann Brunn \cite{Brunn:1892a} en 1892, selon laquelle toute structure connective intègre finie est la structure d'un entrelacs de $\mathbf{R}^3$. Brunn a donné, pour toute structure connective intègre finie, l'idée principale de la construction d'un tel entrelacs, construction fondée sur les \emph{entrelacs brunniens}, et une démonstration complète de la représentabilité par entrelacs de toute structure connective intègre finie a finalement été donnée en 1985 par Kanenobu \cite{Kanenobu:198504, Kanenobu:1986}, après un travail intermédiaire, dans les années 1960, de Debrunner \cite{Debrunner:1964}.

Dans le présent article, nous appellerons \emph{conjectures brunniennes} toute affirmation portant sur la réalisabilité de toute structure connective intègre finie, non plus en termes d'entrelacs, mais par les objets auxquels nous associons une ou plusieurs structures connectives, en l'occurrence : des états quantiques dans la section \ref{section structure etats}, des dispositifs multilocaux
dans la section \ref{section structure relationnelle des dispositifs}, des familles de variables aléatoires dans la section \ref{section dispositifs probabilistes}. 

Par exemple, la conjecture brunnienne pour la structure d'intrication globale d'un état quantique pur%
\footnote{voir la définition \ref{df structures de desintrication}, section \ref{subsubsection definition structures de desintrication de psi} page \pageref{df structures de desintrication}.} 
 consiste à affirmer que pour toute structure connective intègre finie $\kappa$, il existe un état quantique pur dont la structure d'intrication globale est précisément la structure $\kappa$.

\subsection{Engendrement de structures connectives}

Pour tout ensemble $E$, l'ensemble des structures connectives intègres dont $E$ peut être muni constitue un treillis complet pour l'inclusion. La plus fine d'entre elle est appelée \textit{structure discrète}, ou encore, pour éviter tout risque de confusion, \textit{structure discrète intègre}. La moins fine est la structure grossière ou indiscrète.

Comme dans le cas des espaces topologiques, une conséquence de l'organisation en treillis complets des structures connectives (intègres) sur un ensemble $E$ est la notion de structure connective (intègre) la plus fine contenant un ensemble donné quelconque $\mathcal{A}$ de parties de $E$. L'engendrement de cette structure à partir de l'ensemble $\mathcal{A}$ est décrit dans \cite{Dugowson:201012}, §\,2, théorème 3. Une telle construction justifie l'appellation \og structure connective intègre \emph{engendrée} par $\mathcal{A}$\fg\, pour désigner la structure en question. On note \label{notation structure connective engendree} $[\mathcal{A}]_1$ ou $[\mathcal{A}]$ la structure connective intègre engendrée par $\mathcal{A}$.

\subsection{Ordre connectif}\label{subsection ordre connectif}

Dans \cite{Dugowson:201203}, nous avons défini l'ordre connectif  $\Omega(X)$ d'un espace connectif $X$ quelconque. C'est un ordinal, généralement transfini. Dans le cas des espaces connectifs intègre \emph{finis}, l'ordre connectif est un entier naturel, plus simple à définir : l'ordre connectif $\Omega(X)$ d'un espace connectif \emph{fini} intègre $X$ coïncide avec l'ordre connectif défini dans \cite{Dugowson:201012}, à savoir la hauteur du graphe orienté acyclique $G_X$ constitué des connexes irréductibles, muni de la relation d'inclusion, où la notion de connexe irréductible est donnée par la définition suivante.

\begin{df}[Connexes irréductibles]  Soit $X=(\vert X\vert,\kappa_X)$ un espace connectif intègre. Une partie connexe $K\in\kappa_X$ est dite \emph{irréductible} si et seulement si elle n'appartient pas à la structure connective intègre engendrée par les autres parties : $K\notin [ \kappa_X\setminus \{K\} ]_1$. 
\end{df}

Les parties connexes irréductibles sont également appelées les \emph{points génériques} de l'espace connectif considéré.

L'ordre connectif d'un espace connectif $X=(\vert X\vert,\kappa)$ sera également appelé l'ordre connectif de la structure connective $\kappa$, et par conséquent sera également notée $\Omega(\kappa)$.

\section{Quantique : deux formalismes de base}\label{section formalismes quantiques}

Dans cette section \ref{section formalismes quantiques}, pour laquelle nous nous sommes essentiellement appuyé sur l'ouvrage de Chuang et Nielsen,\cite{ChuangNielsen:2010}, nous rappelons les deux formalismes mathématiques, d'ailleurs conceptuellement assez différents, réglant le comportement général des systèmes quantiques : celui, $\mathcal{FV}$, des vecteurs d'état (\ref{subsection formalisme vecteurs}), et celui,  $\mathcal{FD}$, des \emph{opérateurs de densité} (\ref{subsection formalisme densite}), avant de préciser comment passer d'un formalisme à l'autre. 

Nous commencerons toutefois par préciser certaines notions communes aux deux formalismes  concernant les notions de système, d'état et de mesure quantiques, l'utilisation d'espaces de Hilbert, ainsi que la composition de systèmes (\ref{subsection notions communes}). 

\begin{rmq} Certains innovations terminologiques nous ont paru nécessaires par rapport à la terminologie usuelle. Bien entendu, dans ce cas, nous les signalons clairement.
\end{rmq}

\begin{rmq} La présentation usuelle de ces formalismes s'appuie notamment sur des postulats. Notre propre présentation des choses nous conduit parfois à ne présenter que des morceaux de ces postulats, que nous signalons en parlant également de principes : \emph{principe de la mesure projective}, \emph{principe du mixage des états}, etc.
\end{rmq}

\subsection{Notions communes aux deux formalismes}\label{subsection notions communes}

\subsubsection{Notion de système quantique}\label{subsection notion systeme}

\begin{df}\label{df systeme quantique} Un \emph{système quantique} $\mathcal{S}$ est défini, dans un formalisme $\mathcal{F}$, par la donnée 
\begin{itemize}
\item d'un espace de Hilbert $H_\mathcal{S}$, appelé \emph{espace des états} de $\mathcal{S}$,
\item d'un ensemble $\widetilde{\mathcal{S}}_\mathcal{F}$, entièrement déterminé par l'espace $H_\mathcal{S}$, et  dont les éléments constituent les \emph{états}  dans lesquels le système $\mathcal{S}$ est susceptible, \emph{éventuellement}, de se trouver.
\end{itemize}
\end{df}

\begin{rmq}
Pour un système dont l'espace des états est $H_\mathcal{S}$, la définition de l'ensemble $\widetilde{\mathcal{S}}_\mathcal{F}$ dépend du formalisme $\mathcal{F}$ utilisé. Pour les distinguer, nous appellerons parfois \emph{états vectoriels} les états considérés dans le formalisme $\mathcal{FV}$ des vecteurs d'état, et \emph{états de densité} ceux qu'emploie le formalisme $\mathcal{FD}$ des opérateur de densité. En tout cas, il importe de souligner qu'aussi bien avec l'un que l'autre des deux formalismes en question, l'espace de Hilbert $H_\mathcal{S}$ \emph{n'est pas} l'ensemble des états $\widetilde{\mathcal{S}}$ de $\mathcal{S}$. On pourrait regretter, dans ces conditions, que l'espace de Hilbert en question soit couramment appelé \emph{espace des états}. Heureusement, en pratique, cela ne semble pas devoir être source de trop grandes confusions.
\end{rmq}

\begin{rmq}
Le mot \emph{éventuellement} que nous avons écrit dans la définition \ref{df systeme quantique} ci-dessus vient de ce que, dans le cas du formalisme des vecteurs d'état, il peut arriver qu'un système quantique ne se trouve dans aucun état. Par ailleurs, dans le formalisme des opérateurs de densité, nous verrons qu'un système quantique peut se trouver dans plusieurs états différents, selon la connaissance que l'on a de ce système.
\end{rmq}


\begin{rmq} D'une façon générale, l'expression \emph{système quantique} désignera pour nous uniquement des systèmes qui, en dehors des expériences de mesure, restent \emph{fermés}, c'est-à-dire n'échangeant pas d'information, et par conséquent pas non plus d'énergie, avec d'autres systèmes. Il faut toutefois souligner que les corrélations dues à l'intrication quantique rendent problématique, et en tout cas plutôt subtile, cette notion de fermeture. Elle se présente du reste de manière différente dans les deux formalismes.
\end{rmq}

\begin{rmq}
En général, un système quantique connaît une évolution continue de son état \og au cours du temps\fg (quitte à interpréter en termes relativistes la signification de cette expression : temps propre d'une particule, etc...), une telle évolution étant régie par l'équation de Schrödinger ou l'une de ses variantes (Heisenberg, Dirac, etc). Nous ne considérerons pas du tout cet aspect des choses dans le présent article.
Par contre, il peut se produire, en particulier au sein d'un espace-temps relativiste, qu'un système quantique soit créé ou soit détruit à la faveur de la disparition ou de la création d'autres systèmes quantiques (pour distinguer deux systèmes quantiques, on pourra faire en sorte que les espaces de leurs états respectifs soient disjoints).
\end{rmq}

\subsubsection{\`{A} propos des espaces de Hilbert, des bras et des kets}\label{subsubsection Hilbert}

Nous ne considérerons dans le présent article que des espaces de Hilbert \emph{de dimension finie}, le plus souvent une puissance de $2$. Par ailleurs, les espaces de Hilbert considérés seront définis sur le corps des complexes $\mathbf{C}$, bien qu'en pratique nous n'utiliserons que des nombres réels. Le produit scalaire ${<\cdot \,\vert\, \cdot>}$ sur l'espace de Hilbert $H$ sera donc hermitien, et tout opérateur positif --- \emph{i.e.} \og semi-defini positif\fg, autrement dit dont la forme quadratique associée ne prend que des valeurs positives ou nulles --- est automatiquement auto-adjoint. 

L'adjointe (ou conjuguée) d'une matrice (ou d'un endomorphisme) $A$ sera notée $A^\dagger$. Une base orthonormée de $H$ étant donnée, un vecteur $\psi$ de $H$ s'identifie à une matrice colonne, tandis que la forme linéaire $\psi^\dagger$ définie pour tout $\varphi\in H$ par $\psi^\dagger(\varphi)={<\psi,\varphi>}\in \mathbf{C}$ s'identifie à la matrice ligne adjointe de celle de $\psi$. 

Dirac a introduit une notation astucieuse pour désigner les vecteurs de $H$ et leurs adjoints : la notation des $bras$ et des $kets$. Selon cette notation,
les vecteurs $\psi$, $\varphi$, $a$, $b$ etc...  de l'espace $H$ seront également notés sous forme de \emph{kets} :
$\vert \psi>$, 
$\vert\varphi>$, 
$|a>$, $|b>$, ...

\begin{rmq} On utilisera également couramment les notations
$\vert 0>$, $\vert 1>$, où $\vert 0>$ \emph{ne désigne en aucun cas le vecteur nul}, mais un vecteur unitaire que les informaticiens ont de bonnes raisons de vouloir noter ainsi pour le distinguer du vecteur $\vert 1>$, ces deux-là étant orthogonaux.
\end{rmq}

Selon la notation de Dirac, les formes linéaires sur $H$, adjointes des vecteurs, s'écrivent sous forme de \emph{bras} : la forme linéaire adjointe du vecteur $\vert\psi>$ sera ainsi écrite  $<\psi\vert$. De cette manière, l'action de la forme $<\psi\vert$ sur un vecteur $\vert\varphi>$, qu'on aurait pu noter $<\psi\vert\vert\varphi>$, se notera simplement $<\psi\vert\varphi>$, ce qui est bien le produit scalaire attendu.

\begin{rmq}[Sur l'expression $<v\vert A\vert w>$]\label{rmq sur ecriture braket v A w}
Notons provisoirement $(.,.)$ le produit scalaire sur $H$. 
Pour tout endomorphisme $A$ et tous vecteurs $u$ et $v$, on a :
\[<v\vert \vert A\vert w>=((<v\vert)^\dagger,A\vert w>)=(\vert v>, A\vert w>)=(A^\dagger \vert v>,\vert w>),\] d'où
\[<v\vert \vert A\vert w>=((<v \vert A)^\dagger,\vert w>)=<v\vert A\vert\vert w>.\]
Les deux expressions $<v\vert \vert A\vert w>$ et $<v\vert A\vert\vert w>$ désignant dans tous les cas le même scalaire, il est naturel de noter simplement celui-ci $<v\vert A\vert w>$.
\end{rmq}

\subsubsection{Expériences de mesure, observables}

Les états quantiques d'un système ne peuvent pas, en général, être connus expérimentalement. Néanmoins, tout système quantique peut faire l'objet d'expé\-riences de mesure, visant à obtenir au moins des informations partielles sur les états dans lesquels il pourrait se trouver, expé\-riences qui, le plus souvent, modifient d'ailleurs l'état quantique du système. Plus préci\-sément, on dispose pour tout système quantique d'un ensemble de procédures formelles que nous appellerons \emph{expériences de mesure} ou simplement \emph{mesures}, procédures qui indiquent, pour tout état $e$ du système, la probabilité, au terme de cette procédure  : 
\begin{itemize}
\item d'obtenir telle ou telle valeur $\lambda$, appelée \emph{résultat de la mesure}, appartenant à un ensemble de valeurs possibles associé à cette procédure,
\item que le système connaisse une transition vers tel ou tel état $e'$, fonction à la fois de l'état antérieur $e$ du système et du résultat $\lambda$ obtenu.
\end{itemize}

Il existe plusieurs classes plus ou moins générales de telles procédures formelles de mesures. Dans cet article, nous utiliserons uniquement ce qu'on appelle les \emph{mesures projectives}, qui reposent sur la notion d'\emph{observable}\footnote{Nous renvoyons le lecteur intéressé à l'ouvrage \cite{ChuangNielsen:2010} pour la définition plus générale de mesures quantiques fondées sur l'utilisation de \emph{systèmes d'opérateurs de mesure}, les différences entre une mesure faite selon un systèmes d'opérateurs quelconques avec le cas particulier d'une mesure projective sont : premièrement, qu'un système d'opérateurs quelconque  peut conduire à un nombre quelconque de résultats différents pour la mesure effectuée; deuxièmement, que ces résultats ne sont pas néces\-saire\-ment numériques; et enfin, que les états dans lesquels se retrouvent le système juste après la mesure résultent de l'action d'opérateurs quelconques et non néces\-saire\-ment de projections orthogonales.}

\begin{df} Une \emph{observable} $A$ du système quantique $\mathcal{S}$ est un opérateur hermitien\footnote{Un opérateur hermitien est la même chose qu'un opérateur auto-adjoint.} sur   $H=H_\mathcal{S}$, autrement dit un opérateur linéaire diagonalisable avec des valeurs propres réelles et dont les sous-espaces propres sont deux à deux orthogonaux.
\end{df}

\begin{rmq}
La définition donnée ci-dessus d'une observable est difficile à comprendre directement. Elle se comprend mieux à la lumière de la description de la procédure formelle de mesure --- spécifique à chacun des deux formalismes employés --- associée à toute observable. En tout état de cause, une telle définition a un coté un peu artificiel, dans la mesure où deux types de données assez différents s'y trouve pour des raisons de commodité accolés : d'une part des sous-espaces orthogonaux supplémentaires, d'autre part des valeurs réelles attachées à ces sous-espaces.
\end{rmq}

\subsubsection{Systèmes composés}

\begin{df}\label{df systeme compose} Étant donnée une famille $(\mathcal{S}_i)_{i\in I}$ de systèmes quantiques, on appelle système quantique composé par cette famille un système $\mathcal{S}$ dont l'espace des états est le produit tensoriel de celui des $\mathcal{S}_i$ :
\begin{equation}\label{formule espace compose}
H_\mathcal{S}=\bigotimes_{i\in I} H_i,
\end{equation} où on a posé $H_i=H_{\mathcal{S}_i}$.
\end{df}

En outre, lorsque les sous-systèmes $\mathcal{S}_i$ se trouvent dans des états donnés, il en découle l'attribution au système global $\mathcal{S}$ d'un état spécifique, selon des règles qui dépendent du formalisme employé (voir ci-après les sections  \ref{subsubsec vectoriel composition} et \ref{subsubsec densite composition}).

\begin{rmq}\label{rmq brakets pour prod tensoriels hilberts} On rappelle que le produit scalaire sur l'espace $\bigotimes_{i\in I} H_i$ 
est donné par le produit des produits scalaires sur chaque $H_i$, 
de sorte que l'on peut écrire
\[
<\psi_1\psi_2\cdots\psi_k\vert \varphi_1\varphi_2\cdots\varphi_k>=<\psi_1\vert\varphi_1><\psi_2\vert\varphi_2>\cdots <\psi_k\vert\varphi_k>
\]
où la notation {$<bra\vert ket>$} a été étendue aux produit tensoriels en posant
\[
\vert \varphi_1\varphi_2\cdots\varphi_k>=\varphi_1\otimes\varphi_2\otimes \cdots\otimes\varphi_k,
\]
et
\[
<\psi_1\psi_2\cdots\psi_k\vert =\vert \psi_1\psi_2\cdots\psi_k>^\dagger.
\] 
\end{rmq}

\subsection{Formalisme $\mathcal{FV}$ des vecteurs d'état}\label{subsection formalisme vecteurs}

\subsubsection{États (vectoriels) d'un système quantique}

Dans le formalisme des vecteurs d'état, l'ensemble des états (vectoriels)  $\widetilde{\mathcal{S}}_\mathcal{FV}$ d'un système quantique $\mathcal{S}$ est l'ensemble des \emph{droites vectorielles} de l'espace des états ${H_\mathcal{S}}$. Autrement dit, $\widetilde{\mathcal{S}}_\mathcal{FV}$ est l'espace projectif associé à ${H_\mathcal{S}}$.

Par un abus de langage courant, nous parlerons des vecteurs appartenant à l'espace de Hilbert ${H_\mathcal{S}}$ comme constituant eux-mêmes les états du système quantique $\mathcal{S}$, alors que deux vecteurs non nuls colinéaires de cet espace, engendrant la même droite vectorielle, représenteront en fait le même état.

En pratique, on se restreint presque toujours à l'utilisation de vecteurs \emph{unitaires} $\psi$ --- \emph{i.e.} tels que $<\psi\vert \psi>= 1$ --- de l'espace ${H_\mathcal{S}}$. En particulier, un tel vecteur unitaire $\psi\in H_\mathcal{S}$ et son opposé $-\psi$ représenteront \emph{le même} état quantique, de même que tout vecteur unitaire de la forme $e^{i\theta}\psi$.

\subsubsection{Mesures projectives}

\`{A} toute observable $M$ de sous-espaces propres $(E_\lambda)_{\lambda\in \Lambda}$, autrement dit telle que 
\[M=\sum_{\lambda\in\Lambda}\lambda P_\lambda\]
où les $P_\lambda$ désignent les projections orthogonales sur les $E_\lambda$, vérifiant donc
\[\sum_{\lambda\in\Lambda} P_\lambda=Id,\]
se trouve associée la procédure de mesure suivante : le système quantique étant dans un état (généralement inconnu) 
\[\psi=\sum_{\lambda\in\Lambda} P_\lambda(\psi)\]
avec $<\psi\vert\psi>=1$, il y a une probabilité 
\[
p(\lambda)=\parallel P_\lambda(\psi) \parallel^2=<P_\lambda(\psi)\vert P_\lambda(\psi)>=<\psi\vert P_\lambda(\psi)>=<\psi\vert P_\lambda\vert \psi>
\]
que le résultat numérique pour la mesure effectuée soit le nombre réel $\lambda$, 
le système sautant alors de l'état $\psi$ à l'état 
\begin{equation}\label{formule etat arrive}
P_\lambda(\psi)\simeq \dfrac{P_\lambda(\psi)}{\parallel P_\lambda(\psi) \parallel},
\end{equation} où le vecteur unitaire $\dfrac{P_\lambda(\psi)}{\parallel P_\lambda(\psi) \parallel}$  est bien définie si la probabilité en question est non nulle.


D'après le théorème de Pythagore, on a
\[
\sum_{\lambda\in\Lambda} \parallel P_\lambda(\psi) \parallel^2 =1,
\]
autrement dit la somme des probabilités considérées est bien égale à $1$.

\noindent \textbf{Notation}. Pour tout état vectoriel $a\in H$ et toute observable $M=\sum_{\lambda\in\Lambda}\lambda P_\lambda$, nous noterons $M[a]$ l'ensemble des états vectoriels auxquels peut conduire la mesure de $M$ lorsque le système est initialement dans l'état $a$ , autrement dit
\begin{equation}\label{formule notation ensemble des etats apres mesure}
M[a]=\{\dfrac{P_\lambda(a)}{\parallel P_\lambda(a) \parallel}, \lambda\in \Lambda, P_\lambda(a)\neq 0_H\}.
\end{equation}

\begin{rmq} Chaque observable $M$ donnant lieu à une procédure de mesure entièrement spécifiée, on parlera aussi bien de l'\emph{expérience de mesure $M$}, voire de la \emph{mesure $M$}, que de l'\emph{observable $M$}.
\end{rmq}

\begin{exm}[Opérateur d'homothétie]
Si $M$ est l'opérateur (ou la matrice) nulle, ou plus généralement si $M$ est une homothétie --- en particulier l'identité de $H$ --- l'expérience de mesure correspondante est celle qui consiste à décider à l'avance le résultat et à ne rien faire. Avec une probabilité égale à $1$ l'état du système (toujours inconnu) s'en trouve inchangé. Cet exemple jouera un rôle fondamental dans la notion de mesure partielle d'un système intriqué (voir plus loin la définition \ref{df mesure partielle}).
\end{exm}


\subsubsection{Systèmes composés et intrication}\label{subsubsec vectoriel composition}

\paragraph{Principe de composition des états} Un postulat fondamental de la mécanique quantique\footnote{Voir \cite{ChuangNielsen:2010}, p.\, 94.}, stipule que lorsque les systèmes d'une famille $(\mathcal{S}_i)_{i\in I=\{1,...,k\}}$
 sont respectivement dans les états $\psi_1$, ..., $\psi_k$, alors le système $\mathcal{S}$ composé des $\mathcal{S}_i$ se trouve dans l'état
\begin{equation}\label{formule etat vectoriel compose}
\psi= \bigotimes_{i\in I} \psi_i\in \bigotimes_{i\in I} H_{\mathcal{S}_i}=H_\mathcal{S},
\end{equation}
ce qu'en notations \emph{kets} on écrit encore\footnote{Voir la remarque \ref{rmq brakets pour prod tensoriels hilberts}.}
\[\vert \psi>={\vert\psi_1>}\otimes{\vert\psi_2>}\otimes ... \otimes {\vert\psi_k>}=\vert \psi_1\psi_2...\psi_k>.\]

Cependant, le système $\mathcal{S}$ peut aussi se trouver dans des états non factorisables.

\begin{df} Nous dirons qu'un état $\psi\in H_\mathcal{S}=\bigotimes_{i\in I} H_{\mathcal{S}_i}$ est \emph{complètement séparable} s'il est
 complètement factorisable, autrement dit s'il peut s'écrire sous la forme
\[{\vert \psi>}={\vert\psi_1>}\otimes{\vert\psi_2>}\otimes ... \otimes {\vert\psi_k>}.\]
Un état non complètement séparable sera dit \emph{partiellement intriqué}.
En outre, s'il existe une partition\footnote{Voir la note \ref{note partition}.}  $I=I_A\cup I_B$ telle que $\psi\in H_\mathcal{S}$ puisse s'écrire, à l'ordre des facteurs près\footnote{Voir la remarque \ref{rmq ordre des facteurs}.}, sous la forme 
\[\psi=\psi_A\otimes \psi_B,\] avec, pour $j\in\{A,B\}$, \[\psi_j\in\bigotimes_{i\in I_j}H_{\mathcal{S}_i},\] nous dirons que $\psi$ est \emph{partiellement séparable}. Si $\psi$ n'est pas partiellement séparable, nous dirons qu'il est \emph{globalement intriqué}.
\end{df}

\begin{rmq} Le mot \og partiellement\fg\, doit être compris ici au sens large, \emph{i.e.} équivalent à \og partiellement ou complè\-tement\fg. Par conséquent, un état complètement sépa\-rable est également \og partiellement séparable\fg.
\end{rmq}

\begin{rmq}\label{rmq vectoriel differences avec terminologie usuelle} La terminologie usuelle appelle état \emph{séparable} ce que nous appelons un état \emph{complètement séparable}, et par voie de conséquence elle appelle état \emph{intriqué} ce que nous appelons état \emph{partiellement intriqué}. Ces appellations usuelles se comprennent bien dans le cas où il n'y a que deux sous-systèmes, puisqu'il n'y a pas dans ce cas de différence entre un état complétement séparable et un état partiellement séparable. Dès qu'il y  a au moins trois sous-systèmes, les précisions apportées par les adverbes \emph{complètement} et \emph{partiellement} nous ont semblées indispensables. 
\end{rmq}

\begin{rmq}\label{rmq theoreme de decomposition de Schmidt} Le théorème de décomposition de Schmidt affirme que tout état vectoriel unitaire $\psi\in H_A\otimes H_B$ d'un système composé de \emph{deux} sous-systèmes peut s'écrire sous la forme
\[\psi=\sum_r \alpha_r \psi_r^A\otimes \psi_r^B,\] les $(\psi_r^A)_r$ formant une famille orthonormée de $H_A$ et les $(\psi_r^B)_r$  une famille orthonormée de $H_B$, et les $\alpha_r$ étant des réels strictement positifs, l'ensemble des $\alpha_r$ étant d'ailleurs unique, c'est-à-dire entièrement déterminé par la donnée de $\psi$. L'état $\psi$ est alors séparable si et seulement si le cardinal de cet ensemble, c'est-à-dire le nombre de termes de cette somme, appelée \emph{rang de Schmidt de $\psi$}, est égal à $1$. 
\end{rmq}

Une remarque évidente mais qui mérite réflexion est que lorsqu'un système composé se trouve dans un état intriqué alors plusieurs des sous-systèmes --- tous, dans le cas d'un état globalement intriqué --- connaissent une situation curieuse : \textbf{ils ne sont dans aucun état !}. Ou alors, si l'on considère qu'un système quantique devrait, par définition, se trouver toujours dans un état ou un autre, il faudrait décider que de tels sous-systèmes ne sont pas des systèmes. Comme il semble légitime de douter qu'il s'agisse en tout cas de systèmes véritablement fermés, ce point de vue pourrait en effet être soutenu. Pour notre part, en définissant un système par la donnée de son espace (et de son ensemble) d'états mais sans préciser qu'il devait néces\-saire\-ment se trouver toujours dans tel ou tel état, notre conclusion est différente : les sous-systèmes qui composent un système plus vaste restent des systèmes quantiques fermés, mais il faut alors admettre qu'un système quantique peut parfois n'être dans aucun état.

Dans ces conditions, qu'est-ce que faire une mesure quantique sur un sous-système  qui ne se trouve dans aucun état quantique particulier ? La définition suivante répond à cette question. 

\begin{df}[Mesures partielles]\label{df mesure partielle} Étant donnés $(\mathcal{S}_i)_{i\in I=\{1,...,k\}}$ une famille de systèmes quantiques composant un système  $\mathcal{S}$, et étant donnée une observable $M_i$ sur l'espace $H_{\mathcal{S}_i}$, on appelle \emph{expérience de mesure partielle} de $M_i$ l'expérience de mesure sur le système $\mathcal{S}$ de l'observable définie, à l'ordre des facteurs près\footnote{Voir la remarque \ref{rmq ordre des facteurs}.}, par 
\[M_i\otimes Id_{(\bigotimes_{j\neq i} H_j)}=M_i\otimes (\bigotimes_{j\neq i} Id_{H_j}),\] où on a posé $H_j=H_{\mathcal{S}_j}$
.
\end{df}

\subsection{Formalisme $\mathcal{FD}$ des opérateurs de densité}\label{subsection formalisme densite}

Le formalisme quantique en termes d'opérateurs de densité a été introduit par John von Neumann et, indépendamment, par Lev Landau en 1927. Il constitue d'une certaine façon une généralisation probabiliste du formalisme par vecteurs d'état.

\subsubsection{États (de densité) d'un système quantique}

Conformément à la définition \ref{df systeme quantique}, le formalisme des opérateurs de densité attache à tout système quantique $\mathcal{S}$ d'espace des états l'espace de Hilbert $H_\mathcal{S}$ un \emph{ensemble} d'états, mais celui-ci n'est plus celui donné par le formalisme des vecteurs d'état. Afin de distinguer les deux sortes d'états, nous appellerons \emph{états de densité} ceux que considère le formalisme des opérateurs de densité.

Avant de préciser ce que sont les états de densité d'un système quantique, faisons une remarque préliminaire au sujet du statut épistémologique des états en question : celui-ci est quelque peu ambigu, dans la mesure où, dans ce formalisme, l'état de densité d'un système n'est pas une donnée purement objective, mais peut dépendre également de la connaissance que nous avons à son sujet. En fait d'état, ce qui est formalisé ici se rapproche davantage d'\emph{une description probabiliste de l'état} du système, de sorte qu'un système quantique donné pourra se trouver dans plusieurs états de densité différents en fonction des connaissances que nous aurons à son sujet. Nous donnerons ci-après plusieurs exemples de cette situation (voir page \pageref{paragraph mixture}  le \emph{principe de mixage}, ainsi que la remarque \ref{rmq plusieurs etats densite par oubli resultat mesure} et l'exemple \ref{exm plusieurs etats densite par partiel puis recomposition}).

\begin{df}\label{df etats de densite} Étant donné un système quantique  $\mathcal{S}$ d'\emph{espace} des états $H=H_\mathcal{S}$, \emph{l'ensemble des états de densité} de $\mathcal{S}$ est l'ensemble\label{df notation TH} $\widetilde{\mathcal{S}}_\mathcal{FD}=\mathcal{T}(H)$ des opérateurs $\rho\in\mathcal{L}(H)$ positifs et de trace $1$. Autrement dit, un état de densité $\rho$ de $\mathcal{S}$ est un opérateur sur l'espace $H_\mathcal{S}$ vérifiant
\[\forall \psi\in H_\mathcal{S}, <\psi\vert\rho\vert\psi>=\psi^\dagger\rho \psi\geqslant 0,\]
et
\[tr(\rho)=1.\]
\end{df}

\begin{rmq}L'espace $H_\mathcal{S}$ étant défini sur le corps des complexes $\mathbf{C}$, la positivité de $\rho$ équivaut au fait d'être hermitien ($\rho=\rho^\dagger$) avec des valeur propres $\lambda$ (néces\-saire\-ment réelles) toutes positives ou nulles : $\lambda\geqslant 0$.
\end{rmq}

La définition \ref{df etats de densite} ci-dessus constitue une partie du premier postulat\footnote{Voir \cite{ChuangNielsen:2010}, p.\,102.} du formalisme des opérateurs de densité. Elle doit être complétée par l'énoncé du principe de mixage suivant :

\paragraph{Principe de mixage.}\label{paragraph mixture} Si l'on sait que le système $\mathcal{S}$ doit se trouver dans l'un des états $\rho_1$, {...,} $\rho_m$ avec des probabilité respectives $p_1$,{...,} $p_m$ telles que $\sum_{1\leqslant j\leqslant m} p_j=1$, on attribuera au système $\mathcal{S}$ l'état de densité $\rho= \sum_{1\leqslant j\leqslant m} p_j \rho_j$. Ce principe illustre le fait que l'état de densité d'un système dépend de notre connaissance à son sujet.

\begin{df} Un état de densité $\rho$ est dit \emph{pur} lorsqu'il existe $\psi\in H_\mathcal{S}$ tel que
\[\rho=\vert \psi><\psi\vert.\]
Dans le cas contraire, il est dit \emph{mixte}.
\end{df}

En utilisant le fait que, pour tout $\rho\in\mathcal{L}(H)$, la trace de $\rho$ peut s'écrire
\begin{equation}\label{formule trace base ON}
tr(\rho)=\sum_{j=1}^{j=d}<e_j\vert \rho\vert e_j>,
\end{equation} avec $(e_j)_{j\in\{1,..., d\}}$ une base orthonormée quelconque de $H$,
on obtient immédiatement la formule suivante, très utile :
pour tous vecteurs $\varphi$ et $\psi$ d'un espace de Hilbert $H$, on a 
\begin{equation}\label{formule trace de psi phi = phi psi}
<\varphi\vert \psi>=tr(\vert\psi><\varphi\vert).
\end{equation}

On vérifie également sans difficulté la proposition \ref{prop ro pur trace ro2} :
\begin{prop} \label{prop ro pur trace ro2} Pour tout état de densité $\rho$, on a : $tr(\rho^2)\leqslant 1$. 

De plus,
\[\rho\,\mathrm{est}\,\mathrm{pur}\, \Leftrightarrow tr(\rho^2)=1.\]
\end{prop}

\subsubsection{Mesure projective}

Dans cet article, nous ne ferons appel aux procédures de mesure quantique que dans le seul cadre du formalisme des vecteurs d'état, aussi n'indiquons-nous ici qu'à titre indicatif le principe des mesures projectives pour les états de densité. 

\paragraph{Principe de mesure d'une observable.}
Étant donnée une observable $M=M^\dagger=M=\sum_{\lambda\in\Lambda}\lambda P_\lambda$, où les $P_\lambda$ désignent les projections orthogonales sur les sous-espaces propres de $M$, définie sur l'espace de Hilbert $H_\mathcal{S}$ d'un système quantique $\mathcal{S}$ se trouvant dans un état de densité (généralement inconnu) $\rho$, une mesure de cette observable conduit au résultat $\lambda$ avec la probabilité
\[p(\lambda)= tr(P_\lambda \rho),\]
son état de densité étant à cette occasion transformé en 
\[\rho_\lambda=\dfrac{P_\lambda\rho P_\lambda}{tr(P_\lambda \rho)},\] qui est bien défini lorsque cette probabilité est non nulle.

\begin{rmq}[Impact sur l'état d'un système de l'oubli d'une mesure]\label{rmq plusieurs etats densite par oubli resultat mesure} Comme le remarque Chuang et Nielsen (\cite{ChuangNielsen:2010}, page 100), si l'on a effectué une telle mesure mais qu'on en a perdu la trace, on peut toutefois affirmer, d'après le principe de mixage, qu'après cette mesure le système se trouve avec certitude dans l'état 
\[\rho'=\sum_{\lambda\in\Lambda}p(\lambda)\rho_\lambda=\sum_{\lambda\in\Lambda} P_\lambda\rho P_\lambda.\] Si par contre on a noté le résultat $\lambda$ de cette expérience de mesure, et qu'on ne l'a pas perdu, on peut affirmer que le système se trouve alors dans l'état de densité
\[\rho_\lambda=\dfrac{P_\lambda\rho P_\lambda}{tr(P_\lambda \rho)}.\] Ceci illustre qu'un même système peut se trouver en même temps dans deux états de densité différents, selon la connaissance que l'on en a (songer à deux expérimentateurs dont l'un connaîtrait le résultat de l'expérience et l'autre non), et donne à réfléchir sur le statut épistémologique de ces états de densité.
\end{rmq}

\subsubsection{Systèmes composés et intrication}\label{subsubsec densite composition}

Considérons une famille $(\mathcal{S}_i)_{i\in I=\{1,...,k\}}$ de systèmes quantiques, d'espaces respectifs $H_i=H_{\mathcal{S}_i}$, composant un système global $\mathcal{S}$ d'espace $H=H_{\mathcal{S}}=\bigotimes_{i\in I} H_i$.

\paragraph{Principe de composition des états.} Selon l'un des postulats du formalisme des opérateurs de densité\footnote{Voir \cite{ChuangNielsen:2010}, p.\, 102.}, lorsque les systèmes $(\mathcal{S}_i)$ se trouvent dans des états de densité respectifs $\rho_i$, le système global $\mathcal{S}$ se trouve alors dans l'état de densité

\begin{equation}\label{formule etat densite compose}
\rho=\rho_1\otimes \rho_2\otimes...\otimes \rho_k\in \mathcal{T}(H).
\end{equation}

\begin{rmq}\label{rmq def produit tensoriel operateurs} Rappelons que le produit tensoriel de $\rho_A\in\mathcal{L}(H_A)$ et de $\rho_B\in\mathcal{L}(H_B)$ est $\rho_A\otimes \rho_B\in\mathcal{L}(H_A\otimes H_B)$ caractérisé par le fait que pour tout $(\psi_A,\psi_B)\in H_A\times H_B$, on a 
\[
\rho_A\otimes \rho_B(\psi_A\otimes \psi_B)=\rho_A(\psi_A)\otimes \rho_B(\psi_B),
\]
de sorte que, plus généralement, on peut donc écrire
 \[
 \rho_1\otimes \rho_2\otimes...\otimes \rho_k (\vert \psi_1 \psi_2...\psi_k>)=\vert \rho_1(\psi_1)...\rho_k(\psi_k)>.
 \]
\end{rmq}

\begin{df}\label{df correlation et intrication densite} Nous dirons qu'un état global de densité $\rho\in \mathcal{T}(H)$ est  \emph{complètement acorrélé} s'il est de la forme 
\[\rho=\rho_1\otimes \rho_2\otimes...\otimes \rho_k\in \mathcal{T}(H).\] S'il existe une partition\footnote{Voir la note \ref{note partition}.} $I=I_A\cup I_B$ telle que $\rho$ puisse s'écrire, à l'ordre des facteurs près\footnote{Voir la remarque \ref{rmq ordre des facteurs}.}, sous la forme
\[\rho=\rho_A\otimes \rho_B,\] avec $\rho_A\in \mathcal{T}(\bigotimes_{i\in I_A} H_i)$ et $\rho_B\in \mathcal{T}(\bigotimes_{i\in I_B} H_i)$, nous dirons que $\rho$ est \emph{partiellement acorrélé}.
Un état de densité  non complètement acorrélé sera dit \emph{partiellement corrélé}, et un état non partiellement acorrélé sera dit \emph{complétement corrélé}.

En outre, $\rho$ sera dit  \emph{complètement séparable} s'il peut s'écrire sous la forme 
\[\rho=\sum_{r=1}^{r=n} p_r \rho_{r,1}\otimes \rho_{r,2}\otimes...\otimes \rho_{r,k}\in \mathcal{T}(H),\]avec les $p_r\geqslant 0$ des probabilités telles que $\sum_{r=1}^{r=n} p_r=1$.
S'il existe une partition $I=I_A\cup I_B$ telle que $\rho$ puisse s'écrire, à l'ordre des facteurs près, sous la forme
\[\rho=\sum_{r=1}^{r=n}p_r \rho_{r,A}\otimes \rho_{r,B},\]  avec, pour tout $r$,  $\rho_{r,A}\in \mathcal{T}(\bigotimes_{i\in I_A} H_i)$ et $\rho_{r,B}\in \mathcal{T}(\bigotimes_{i\in I_B} H_i)$, nous dirons que $\rho$ est \emph{partiellement séparable}.
Un état de densité non complètement séparable sera dit \emph{partiellement intriqué}, et un état non partiellement séparable sera dit \emph{complètement intriqué}.
\end{df}

\begin{rmq} L'usage du mot \emph{partiellement} dans la définition ci-dessus doit être compris au sens large : par exemple, un état complètement corrélé est \emph{a fortiori} partiellement corrélé, etc.
\end{rmq}

\begin{rmq}\label{rmq densite differences avec terminologie usuelle} La terminologie usuelle appelle
\begin{itemize}
\item \emph{corrélés} les états que nous appelons partiellement corrélés,
\item \emph{séparables} ceux que nous appelons complètement séparables,
\item \emph{intriqués} ceux que nous appelons partiellement intriqués.
\end{itemize}

Ces différences terminologiques, qui visent à davantage de précision, sont cohérentes avec celles déjà soulignées, dans le cas du formalisme vectoriel,  par la remarque \ref{rmq vectoriel differences avec terminologie usuelle}.
\end{rmq}


Quel que soit l'état global de densité $\rho$ dans lequel le système composé $\mathcal{S}$ se trouve, il existe une façon naturelle --- qui n'est de ce fait pas considérée comme faisant partie des postulats de la mécanique quantique --- d'attribuer à chacun des sous-systèmes ${\mathcal{S}_i}$ un état de densité spécifique dont l'expression en fonction de $\rho$ fait appel à la notion de trace partielle, ainsi définie :

\begin{df}[Traces partielles]\label{df trace partielle} Étant donné un espace de Hilbert $H$ et $H=H_A\otimes H_B$ une factorisation de $H$. On appelle trace partielle sur $H_B$ l'unique application linéaire $tr_B:\mathcal{L}(H)\rightarrow \mathcal{L}(H_A)$ qui, à tout opérateur $\rho$ de la forme
\[
\rho=\vert \varphi_A\varphi_B><\psi_A\psi_B\vert=\vert \varphi_A><\psi_A\vert\otimes \vert \varphi_B><\psi_B\vert
\]
associe l'opérateur $tr_B(\rho)\in\mathcal{L}(H_A)$ donné par
\begin{equation}\label{formule trace reduite}
tr_B(\rho)=<\psi_B\vert\varphi_B> \vert \varphi_A><\psi_A\vert. 
\end{equation}
\end{df}


\begin{df}\label{df reduction densite} d'un état de densité à un sous-système] Étant donné $\rho\in T(H)$ un état de densité du système $\mathcal{S}$ composé par la famille $(\mathcal{S}_i)_{i\in I=\{1,...,k\}}$, où $H=H_\mathcal{S}=\bigotimes_{i\in I} H_i$ et $H_i=H_{\mathcal{S}_i}$, et étant donné $j\in I$, on appelle \emph{réduction de $\rho$ à $H_j$} l'opérateur de densité réduit $\rho^j$ défini par
\[
\rho^j=tr_B(\rho),
\]
où $B=\bigotimes_{i\neq j} H_i$.
Plus généralement, pour toute partie non vide $J\subset I$, on appelle \emph{réduction de $\rho$ à $H_J=\bigotimes_{j\in J} H_j$} l'opérateur $\rho^J\in\mathcal{T}(H_J)$ défini par
\[
\rho^J=tr_B(\rho),
\]
où $B=\bigotimes_{i\in \neg J} H_i$, avec $\neg J= I\setminus J$.
\end{df} 

\paragraph{État de densité d'un sous-système} Dans le formalisme $\mathcal{FD}$, il est dit %
\footnote{Voir Chuang et Nielsen \cite{ChuangNielsen:2010}, pages 105-106.}%
que si le système composé $\mathcal{S}$ est dans l'état de densité $\rho$, alors le sous-système $\mathcal{S}_j$ est dans l'état de densité $\rho^j$ obtenu par réduction de $\rho$ à $H_j$.

\begin{exm}[Deux états pour un même système]\label{exm plusieurs etats densite par partiel puis recomposition} Considérons le système $\mathcal{S}$ composé de deux sous-systèmes $\mathcal{S}_A$
 et $\mathcal{S}_B$, avec $H_A=H_{\mathcal{S}_A}=vect\{{\vert 0_A>},{\vert 1_A>}\}$ et $H_B=H_{\mathcal{S}_B}=vect\{{\vert 0_B>},{\vert 1_B>}\}$, et supposons que $\mathcal{S}$ se trouve dans l'état pur $\rho=\vert\psi><\psi\vert$, avec $\psi=\dfrac{\vert 0_A 0_B>+\vert 1_A 1_B>}{\sqrt{2}}$, que nous noterons plus simplement
 \[\psi=\dfrac{\vert 00>+\vert 11>}{\sqrt{2}}.\]

Dans la base $(\vert 00>, \vert 01>, \vert 10>, \vert 11>)$ de $H$, l'opérateur $\rho$ s'écrit matriciellement
\[\mathrm{Mat}_\rho=\dfrac{1}{2}\left(\begin{array}{cccc}
1 & 0 & 0 & 1 \\ 
0 & 0 & 0 & 0 \\ 
0 & 0 & 0 & 0 \\ 
1 & 0 & 0 & 1
\end{array}  \right),\] et la formule (\ref{formule trace reduite}) permet de trouver facilement l'expression matricielle de $\rho^A$ et $\rho^B$ dans les bases respectives $({\vert 0_A>},{\vert 1_A>)}$ et $({\vert 0_B>},{\vert 1_B>)}$ de $H_A$ et $H_B$ :
\[\mathrm{Mat}_{\rho^A}=\mathrm{Mat}_{\rho^B}=\dfrac{1}{2}\left(\begin{array}{cc}
1 & 0 \\ 
0 & 1
\end{array} \right).\] Appliquant le principe de composition des états, on obtient alors pour le système $\mathcal{S}$ entier un état de densité $\rho'=\rho^A\otimes \rho^B$ de matrice
\[\mathrm{Mat}_{\rho'}=\dfrac{1}{4}\left(\begin{array}{cccc}
1 & 0 & 0 & 0 \\ 
0 & 1 & 0 & 0 \\ 
0 & 0 & 1 & 0 \\ 
0 & 0 & 0 & 1
\end{array}  \right).\] On a donc $\rho\neq \rho'$. Cet exemple illustre à nouveau qu'en fonction de la connaissance que l'on a d'un système, celui-ci peut se voir attribuer des états de densité différents. En l'occurrence, la donnée des états mixtes $\rho^A$ et $\rho^B$ dans lesquels se trouvent les sous-systèmes $A$ et $B$ contient moins d'information que celle contenue dans la donnée de l'état intriqué $\rho$ --- l'information perdue est celle relative à la manière dont $A$ et $B$ sont intriqués --- il est donc logique qu'on ne puisse à partir de $\rho^A$ et $\rho^B$ retrouver l'état $\rho$.
\end{exm}

\subsection{Passage d'un formalisme à l'autre}

\subsubsection{Des vecteurs d'état aux opérateurs de densité}

On passe du formalisme des vecteurs d'états à celui des opérateurs de densité en identifiant tout état vectoriel à un état de densité pur, grâce à l'injection
\[
\begin{array}{c}
\widetilde{\mathcal{S}}_\mathcal{FV} \hookrightarrow \mathcal{T}(H)=\widetilde{\mathcal{S}}_\mathcal{FD},\\ 
\psi=\vert\psi> \mapsto \rho= \vert\psi><\psi\vert.
\end{array} 
\]

Cette injection montre que l'espace des états de densité est plus large que celui des états vectoriels. Les états de densité $\rho$ qui peuvent se mettre sous la forme $\vert \psi><\psi\vert$ sont précisément les états purs, et tout naturellement on appelle également états purs les états vectoriels $\psi$ eux-mêmes.

Plus généralement, à toute famille pondérée d'états vectoriels unitaires $(\psi_r,p_r)_{r\in R}$, où les $p_r\geqslant 0$ vérifient $\sum_r p_r=1$, on associe l'état de densité mixte
\[
\rho=\sum_{r\in R
}
p_r \vert\psi_r>
<\psi_r\vert.
\]

%

\subsubsection{Des opérateurs de densité aux vecteurs d'état}

Un état de densité $\rho$ peut toujours être interprété en termes de famille d'états purs pondérée par des probabilités   $(\psi_r,p_r)_{r\in R}$, autrement dit en termes d'états mixtes $\rho=\sum_{r} p_r \vert\psi_r><\psi_r\vert$, ne serait-ce qu'en prenant pour vecteurs $\psi_r$ une base de vecteurs propres de l'opérateur hermitien $\rho$. Néanmoins, la portée d'une telle interprétation est restreinte par le fait que, très souvent, $\rho$ peut s'écrire \emph{de plusieurs manières} sous cette forme.

\section{Structures connectives des états quantiques intriqués}\label{section structure etats}
\sectionmark{Structures connectives des états intriqués}

Padmanabhan K. Aravind, professeur de physique au \emph{Worcester Polytechnic Institute} (Massachusetts),  dans un article \cite{Aravind:1997} publié en 1997 comme chapitre du livre \cite{Shimony:1997}, et Ayumu Sugita, du département de physique appliquée de la \emph{Osaka City University}, dans un article \cite{Sugita:2007} publié en 2007 sur ArXiv, cherchent à comparer intrication quantique et entrelacs dans $\mathbf{R}^3$, et cela essentiellement dans le cas de trois composantes --- bien qu'Aravind considère aussi l'intrication \emph{borroméenne généralisée} à un nombre quelconque de composantes, autrement dit ce qu'on appelle aussi depuis 1976 avec \cite{Rolfsen:1976} les entrelacs \emph{brunniens}\footnote{Sur l'histoire des structures connectives, et en particulier la contribution fondamentale de Brunn à ce sujet, voir \cite{Dugowson:201012} ainsi que le rappel historique figurant dans l'introduction  de  \cite{Dugowson:201306}.} --- avec en particulier la question de savoir si le système $GHZ$ constitué de trois qubits intriqués dans l'état
\begin{equation}\label{formule etat GHZ}
GHZ=\dfrac{\vert 000>+\vert 111>}{\sqrt{2}}
\end{equation}
peut ou non être dit borroméen.

Or, dans ce type de question, le seul aspect des entrelacs qui intervient réellement est leur structure connective\footnote{Sur la structure connective des entrelacs, voir \cite{Dugowson:201012}.}. C'est du reste une question très naturelle, lorsqu'on considère l'intrication quantique en ayant à l'esprit le point de vue connectif, de se demander s'il est possible d'associer de façon naturelle une ou plusieurs structures connectives à tout état d'un système quantique composé. C'est l'objet de la présente partie que de proposer la définition de telles structures, en nous appuyant d'abord sur l'idée développée par Aravind, et ensuite sur celle de Sugita, la principale différence entre les deux points de vue étant qu'Aravind utilise essentiellement le formalisme des vecteurs d'état, tandis que Sugita utilise celui des opérateurs de densité.

\`{A} la fin de la présente section \ref{section structure etats}, nous introduisons la notion d'ordre connectif d'un état quantique intriqué.

\subsection{Structures connectives de désintrication}\label{subsection structures desintrication}

\subsubsection{L'analogie d'Aravind entre systèmes intriqués et entrelacs}

Dans le cas d'un entrelacs, la considération de sous-entrelacs est immédiate : il suffit d'oublier les composantes qui n'en font pas partie. Or, un tel oubli n'a pas d'équivalent évident s'agissant des systèmes quantiques composés, du moins dans le formalisme des vecteurs d'état. Par contre, l'opération qui consiste à obtenir un sous-entrelacs d'un entrelacs donné  comme  résultat d'opérations de \emph{coupures} sur les composantes qui ne lui appartiennent pas a de manière assez évidente un analogue dans le formalisme quantique des états purs, à savoir la transformation%
\footnote{Transformation au moins partiellement désintricante, s'agissant en tout cas de la particule sur laquelle la mesure est effectuée. Curieusement, la désintrication quantique semble être une notion encore relativement peu considérée en tant que telle. Remarquons en outre qu'elle soulève de façon assez aigüe le problème de comprendre l'articulation entre non-localité et contraintes relativistes, dans la mesure où, comme nous le soulignons dans la remarque \ref{rmq chronologie fictive des mesures et des desintrication} page \pageref{rmq chronologie fictive des mesures et des desintrication}, il est impossible de localiser en tant qu'unique événement de l'espace-temps relativiste une telle désintrication : deux mesures constituant des événements séparés par un intervalle de genre espace effectuées sur deux parties intriquées d'un même système quantique peuvent aussi bien prétendre l'une que l'autre être \og à l'origine\fg\, de la désintrication : tout se passe comme si elles ne faisaient qu'actualiser une désintrication dont la cause n'est pas localisable, désintrication qui se produit dès qu'au moins une des deux mesures a lieu, mais qui ne se serait pas produite si aucune des deux mesures n'avaient été faite... } 
des états consécutive à une mesure quantique effectuée sur l'une des particules formant un système intriqué. C'est là l'idée centrale adoptée par Padmanabhan K. Aravind \cite{Aravind:1997}. Le problème est alors le suivant : une même mesure peut parfois conduire, pour le système constitué des autres particules que celle sur laquelle la mesure (partielle) a été effectuée, à un système intriqué ou non. De plus, cela peut également dépendre du type de mesure effectué, un peu comme si dans un entrelacs, selon que l'on effectue sur l'un des n\oe uds qui le compose une coupure de tel ou tel type, les composantes restantes devaient former des entrelacs de structures différentes. C'est en particulier le cas pour l'exemple fondamental autour duquel l'article d'Aravind est construit, l'état $GHZ$  qui 
se révèle conduire soit à une structure borroméenne, soit à une structure totalement connectée selon les expériences de mesure considérées.

\subsubsection{Un éventail de structures connectives}

Il découle de qui précède qu'il y pourrait y avoir plusieurs structures connectives légitimes à envisager pour un même état quantique intriqué. Dans ce qui suit, nous allons effectivement définir, à la section \ref{subsubsection definition structures de desintrication de psi}, tout un éventail de structures connectives pour un état pur d'un système quantique composé d'une famille finie $(\mathcal{S}_i)_{i\in I=\{1,...,k\}}$ de sous-systèmes, chacun d'eux ayant pour espaces d'états un espace de Hilbert $H_i$. Cet état sera 
représenté par un vecteur unitaire $\psi$ :
\[
\psi\in H=\bigotimes_{i\in I} H_i.
\]

Pour cela, nous avons d'abord besoin de donner quelques précisions de vocabulaire et de notations relatives aux ensembles d'états à considérer, à la séparabilité de ces états, aux ensembles de mesures qui peuvent être effectués sur ces états, ainsi que des précisions concernant le quantificateur logique que nous noterons $\exists^\bullet$ pour exprimer le fait qu'une propriété est satisfaite par \emph{certains} éléments, mais \emph{pas tous}.

\subsubsection{$J$-états}

Pour toute partie non vide $J$ de $I$, nous noterons $H_J$ l'espace des états du sous-système $\mathcal{S}_J$ de $\mathcal{S}$, espace défini par
\begin{equation}\label{notation HJ}H_J=\bigotimes_{j\in J} H_j,\end{equation} 
et nous appellerons \emph{état sur $J$} ou \emph{$J$-état} tout état $\varphi$ de $\mathcal{S}_J$, c'est-à-dire tout état défini par un vecteur unitaire $\varphi\in H_J$.

\begin{df} Soit $J\subset I$, avec $card(J)\geqslant 2$, et $J=J_1\cup J_2$ une partition%
\footnote{\label{note partition} Rappelons qu'une partition d'un ensemble est un recouvrement de cet ensemble par des parties disjointes \emph{toutes non vides}.}
de $J$. On dit qu'un $J$-état $\varphi\in H_J$ est \emph{$(J_1,J_2)$-séparable} s'il existe $\varphi_1\in H_{J_1}$ et $\varphi_2\in H_{J_2}$ tels que, \emph{à l'ordre des facteurs près}, on ait
\[\varphi=\varphi_1\otimes \varphi_2.\] 
On note\label{notation etats J1J2 separables} $\mathcal{D}_{(J_1,J_2)}$ l'ensemble des $J$-états $(J_1,J_2)$-séparables.
\end{df}

\begin{rmq}[\og \`{A} l'ordre des facteurs près\fg]\label{rmq ordre des facteurs} Le produit tensoriel n'étant pas commutatif, nous serons souvent conduit à préciser que les formules écrites doivent être comprises \og à l'ordre des facteurs près\fg. Par exemple, dans la formule écrite ci-dessus, $\varphi=\varphi_1\otimes \varphi_2$, cela ne signifie pas seulement que l'on pourrait avoir aussi $\varphi=\varphi_2\otimes \varphi_1$, mais surtout que l'ordre de \emph{tous} les facteurs figurant dans \emph{toute} expression de $\varphi_1$ et de $\varphi_2$ doit éventuellement être rétabli pour entrer dans la définition usuelle des éléments de $\bigotimes_{j\in J} H_j$. Au fond, tout cela est simplement équivalent à l'utilisation d'une version commutative du produit tensoriel définie, pour éviter toute ambiguïté, sur des \emph{copies deux à deux disjointes} des espaces $H_i$. Ceci précisé, si d'aventure il nous arrivait d'oublier de préciser la mention \og à l'ordre des facteurs près\fg, le lecteur ne manquerait pas de rectifier par lui-même.
\end{rmq}

\begin{df}\label{df J-etats separables} $J$ désignant une partie de $I$ non vide et non réduite à un point,
un $J$-état $\varphi\in H_J$ est dit \emph{partiellement séparable} s'il existe une partition $J=J_1\cup J_2$ telle que $\varphi$ soit $(J_1,J_2)$-séparable. On note $\mathcal{D}_{J}$ (ou simplement $\mathcal{D}$ s'il n'y a pas d'ambiguïté) l'ensemble des $J$-états partiellement séparables. Un $J$-état non partiellement séparable sera dit \emph{globalement intriqué}.
\end{df}

Nous noterons en outre $\neg \mathcal{D}_J$ (ou simplement $\neg \mathcal{D}$ s'il n'y a pas d'ambiguïté) l'ensemble des $J$-états globalement intriqués, autrement dit le complémentaire de $\mathcal{D}_J$ dans $H_J$ :
\begin{equation}\label{notation neg mathcal DJ}\neg \mathcal{D}_J = H_J\setminus \mathcal{D}_J.\end{equation}

\begin{rmq} La question de savoir s'il faut élargir la définition \ref{df J-etats separables} aux parties de $I$ réduites à un point n'est pas vraiment importante, dans la mesure où les structures connectives qui seront définies sur la base de ces considérations seront de toute façon intègres, autrement dit les singletons seront considérés dans tous les cas comme connexes.
\end{rmq}

\subsubsection{Expériences déterminantes et projections sur $J$}

Pour toute partie $J\subset I$, on note\label{notation negJ} $\neg J$ le complémentaire de $J$ dans $I$.

\begin{df} Étant donné $L\subset I$, on appelle \emph{expérience déterminante sur $L$} toute expérience de mesure d'une observable $M$ sur le système $\mathcal{S}$ qui, \emph{à l'ordre des facteurs près}\footnote{Voir la remarque \ref{rmq ordre des facteurs}.}, est de la forme 
\[M=(\bigotimes_{l\in L}M_l)\otimes (\bigotimes_{j\in \neg L}Id_j)=(\bigotimes_{l\in L}M_l)\otimes Id_{H_{\neg L}},\]  où, pour tout $l\in L$, $M_l$ désigne un opérateur hermitien de l'espace $H_l$ dont toutes les valeurs propres soient distinctes, \emph{i.e.} dont tous les sous-espaces propres soient de dimension $1$. 
\end{df}

On note \label{notation mathcalM L} $\mathcal{M}_L$ l'ensemble des expériences déterminantes sur $L$. 

\begin{rmq}\`{A} noter que l'observable $\bigotimes_{l\in L}M_l$ peut s'interpréter comme synthé\-tisant la suite des observables $(M_l)_{l\in L}$. Voir à ce sujet la remarque \ref{rmq chronologie fictive des mesures et des desintrication} page \pageref{rmq chronologie fictive des mesures et des desintrication}.
\end{rmq}

\begin{rmq} L'ensemble $I$ ayant été supposé non vide, les ensembles  $\mathcal{M}_L$ ne sont jamais vides. En particulier, pour $L=\emptyset\subset I$, on a $\mathcal{M}_L=\{Id_H\}$. 
\end{rmq}

Soit maintenant $J$ une partie propre de $I$ et $M\in\mathcal{M}_{\neg J}$ une expérience déterminante sur $\neg J$. Les sous-espaces propres des $M_l$ qui composent $M$ étant tous de dimension $1$, 
tout état%
\footnote{Voir la notation donnée par la formule (\ref{formule notation ensemble des etats apres mesure}) page \pageref{formule notation ensemble des etats apres mesure}.} 
$b\in M[\psi]$ auquel la mesure de $M$ peut conduire à partir de l'état $\psi$ est $(J,\neg J)$-séparable, autrement dit peut se factoriser sous la forme 
\[b=(\bigotimes_{l\in L} b_l)\otimes b_{J},\] avec, pour tout $l\in L$, $b_l$ un vecteur propre unitaire de $M_l$, et $b_{J}\in H_J$, une telle factorisation étant unique (à des coefficients multiplicatifs de module $1$ près).  Par conséquent, tout état $b\in M[\psi]$ définit un unique état\label{notation bJ}  $b_{J}\in H_J$, que nous appellerons la \emph{projection de $b$ sur $J$}.

Pour tout $M\in\mathcal{M}_{\neg J}$, on posera en outre
\begin{equation} \label{notation MJa} M_J[\psi]=\{b_J\in H_J, b\in M[\psi]\}.\end{equation}

\begin{rmq} Pour $J=I$, on a $\mathcal{M}_{\neg J}=\{Id_H\}$, et on pose naturellement ${(Id_H)}_J[\psi]=\{\psi\}$.
\end{rmq}

\begin{rmq} Si on n'avait pas imposé que les valeurs propres de chaque $M_l$ soient distinctes, on aurait effectivement pu rencontrer des situations telles que celles-ci : si $u$ et $v$ sont des vecteurs propres orthogonaux de $M_1$ associés à une même valeur propre, une mesure de $M_1$ sur un état intriqué de la forme $\dfrac{\vert u00> + \vert v11>}{\sqrt{2}}$ peut conduire à le laissé inchangé, donc non factorisable.
\end{rmq}

\subsubsection{Le quantificateur $\exists^\bullet$}

Nous définissons le quantificateur \emph{certains mais pas tous}, noté $\exists^\bullet$, par l'équivalence logique suivante, écrite pour toute propriété $P(x)$ dépendant d'une variable $x\in E$ :
\begin{equation}\label{formule quantificateur pas tous}
\exists^\bullet x\in E, P(x) \Leftrightarrow (\exists x\in E, P(x))\,\mathrm{et}\,(\exists x\in E, \neg P(x)).
\end{equation}

\subsubsection{Types d'intrication de $\psi$ sur $J\subset I$}\label{subsubsection intrications J}

Dans toute cette section \ref{subsubsection intrications J},  $J$ désigne une partie de $I$ ayant au moins deux éléments, et représente ainsi un sous-ensemble de l'ensemble des particules constituant le système $\mathcal{S}$.

Rappelons par ailleurs que $\psi$ désigne un état vectoriel du système quantique considéré, autrement dit $\psi\in \bigotimes_{i\in I} H_i$, avec $\psi$ unitaire. 

\begin{df} On dit que l'état $\psi$ est \emph{globalement intriqué sur $J$} si 
\[
\forall M\in \mathcal{M}_{\neg J}, M_J[\psi]\subset\neg\mathcal{D}_J.
\]
\end{df}

Autrement dit, l'état $\psi$ du système est globalement intriqué sur  $J\subset I$  lorsque toute expérience déterminante effectuée sur les autres particules que celles indexées par $J$ conduit néces\-saire\-ment ce sous-système à un état globalement intriqué.

\begin{rmq}\label{rmq I=J} L'état $\psi$  est globalement intriqué sur $I$ lui-même si et seulement si $\psi\in \neg\mathcal{D}$, c'est-à-dire si l'état $\psi$ est intriqué.
\end{rmq}

\begin{df} On dit que $\psi$ est \emph{mélangé}, ou qu'il est \emph{semi-intriqué}, ou encore qu'il est \emph{semi-séparable} sur $J$ si et seulement si 
\[\exists (M,P)\in (\mathcal{M}_{\neg J})^2, \exists (\varphi,\phi)\in M_J[\psi]\times P_J[\psi],(\varphi\in \neg \mathcal{D}_J)\,\mathrm{et}\, (\phi\in\mathcal{D}_J).\]
\end{df}

Autrement dit, lorsque $\psi$ est mélangé sur $J$, le résultat sur $J$ d'une expérience de mesure faite sur les autres sous-systèmes peut, selon les cas, conduire aussi bien à un état globalement intriqué sur $J$ qu'à un état partiellement séparable sur $J$.

\begin{df} On dit que $\psi$ est \emph{totalement mélangé} sur $J$ si et seulement si 
\[\forall M\in \mathcal{M}_{\neg J}, \exists^\bullet \varphi\in M_J[\psi], \varphi\in \mathcal{D}_J.\]
\end{df}

Ainsi, $\psi$  est totalement mélangé lorsque \emph{toute} expérience déterminante effectuée sur les autres sous-systèmes conduit parfois à un état sur $J$ partiellement séparable, et parfois à un état globalement intriqué.

\begin{df} On dit que $\psi$ est \emph{partiellement bien intriqué} sur $J$ si
\[\exists^\bullet M\in \mathcal{M}_{\neg J}, M_J[\psi] \subset \neg \mathcal{D}_J.\]
\end{df}

Autrement dit, $\psi$ est partiellement bien intriqué sur $J$ s'il y est mélangé mais qu'il y a au moins une expérience de mesure sur les autres sous-systèmes conduisant, selon toutes probabilités, à un état globalement intriqué sur $J$. Dans ce cas, $\psi$ n'est donc pas totalement mélangé sur $J$.

\begin{df} On dit que $\psi$ est \emph{partiellement bien séparable} sur $J$ si
\[\exists^\bullet M\in \mathcal{M}_{\neg J}, M_J[\psi] \subset \mathcal{D}_J.\]
\end{df}

Autrement dit, $\psi$ est partiellement bien séparable sur $J$ s'il est mélangé sur $J$ mais qu'il y a au moins une expérience de mesure sur les autres sous-systèmes conduisant, selon toutes probabilités, à un état partiellement séparable sur $J$. Dans ce cas non plus $\psi$ n'est pas totalement mélangé sur $J$.

\begin{df} On dit que $\psi$ est \emph{globalement séparable sur $J$} si
\[
\forall M\in \mathcal{M}_{\neg J}, M_J[\psi] \subset \mathcal{D}_J.
\]
\end{df}

\begin{df} On dit que $\psi$ est \emph{clairement séparable sur $J$} s'il existe une partition de $J$ en deux parties non vides disjointes $J=J_1\cup J_2$ telles que
\[
\forall M\in \mathcal{M}_{\neg J}, M_J[\psi] \subset \mathcal{D}_{(J_1,J_2)}.
\]
\end{df}

Alors que le fait, pour $\psi$, d'être globalement séparable sur $J$ conduit à des états qui peuvent se factoriser différemment selon les expériences de mesure réalisées et, pour chaque expérience, selon les résultats obtenus, la claire séparabilité demande une possibilité de factorisation selon la même partition pour tous les états ainsi obtenus.

\begin{df} On dit que $\psi$ est \emph{totalement séparé sur $J$} si, pour toute partition de $J$ en deux parties non vides disjointes $J=J_1\cup J_2$, on a
\[
\forall M\in \mathcal{M}_{\neg J}, M_J[\psi] \subset \mathcal{D}_{(J_1,J_2)}.
\]
\end{df}

On vérifie facilement que les définitions précédentes conduisent à classer la situation d'intrication de $\psi$ sur $J$ dans l'une des huit situations mutuellement incompatibles suivantes : $\psi$ y est soit globalement intriqué, soit totalement mélangé, soit partiellement bien intriqué mais non partiellement bien séparable, soit partiellement bien séparable mais non partiellement bien intriqué, soit à la fois partiellement bien intriqué et partiellement bien séparable, soit globalement séparable mais non clairement séparable, soit clairement séparable mais non totalement séparé, soit enfin totalement séparé. 

\begin{figure} 
\begin{center}
\includegraphics[bb=0 0 1022 675,width=9cm]{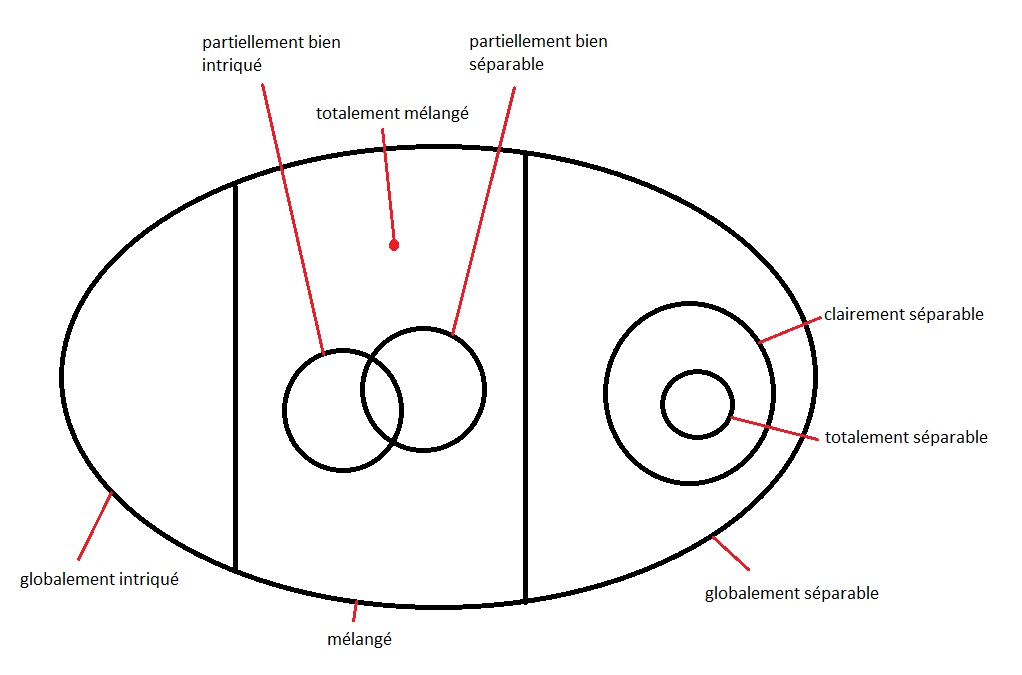}
\caption{Types d'intrication d'un état vectoriel $\psi$ sur une partie $J\subset I$}
\label{figure types intrication de psi sur J}
\end{center}
\end{figure}

Cette classification résulte immédiatement de la proposition suivante, elle-même facile à vérifier.

\begin{prop} Sur $J\subset I$, $\psi$ est  soit globalement intriqué, soit mélangé, soit globalement séparable.
Si $\psi$ est mélangé sur $J$, il y est 
\begin{itemize}
\item soit totalement mélangé,
\item soit partiellement bien intriqué sans être partiellement bien séparable,
\item soit partiellement bien séparable sans être partiellement bien intriqué,
\item soit à la fois partiellement bien intriqué et partiellement bien séparable.
\end{itemize}
Enfin, on a les implications :
$\psi$ totalement séparé sur $J$ $\Rightarrow$ $\psi$ clairement séparable sur $J$ $\Rightarrow$ $\psi$ globalement séparable sur $J$.
\end{prop}

\subsubsection{Structures de désintrication de $\psi$}\label{subsubsection definition structures de desintrication de psi}

Sur la base de la classification précédente, nous pouvons maintenant associer à l'état $\psi$ différentes structures connectives, que nous appellerons les structures de désintrication de $\psi$.

Dans ce but, notant  $\mathcal{Q}(I)$ l'ensemble des parties $J\subset I$ telles que $card(J)\geqslant 2$, nous introduisons les ensembles suivants de parties de $I$ :

\begin{itemize}
\item $GI(\psi)$ : ensemble des $J\in\mathcal{Q}(I)$ telles que $\psi$ est globalement intriqué sur $J$,
\item $BIP(\psi)$ : ensemble des $J\in\mathcal{Q}(I)$ telles que $\psi$ est globalement intriqué ou partiellement bien intriqué sur $J$,
\item $MT(\psi)$ : ensemble des $J\in\mathcal{Q}(I)$ telles que $\psi$ est globalement intriqué ou totalement mélangé sur $J$,
\item $IP(\psi)$ : ensemble des $J\in\mathcal{Q}(I)$ telles que $\psi$ est globalement intriqué ou partiellement bien intriqué ou totalement mélangé sur $J$,
\item $ML(\psi)$ : ensemble des $J\in\mathcal{Q}(I)$ telles que $\psi$ est globalement intriqué ou mélangé sur $J$,
\item $NCS(\psi)$ : ensemble des $J\in\mathcal{Q}(I)$ telles que $\psi$ n'est pas clairement séparable sur $J$.
\end{itemize}

\begin{df}[Structures de désintrications de l'état $\psi$]\label{df structures de desintrication}
On définit sur l'ensemble $I$ les structures connectives intègres suivantes, associées à l'état quantique global $\psi\in \bigotimes_{i\in I} H_i$ :
\begin{itemize}
\item structure d'intrication globale%
\footnote{Rappelons que la notation $[\mathcal{A}]_1$ désigne la structure connective intègre engendrée par $\mathcal{A}$ : voir la section \ref{notation structure connective engendree}, page \pageref{notation structure connective engendree}.}%
 : $\kappa_{GI}(\psi)=[GI(\psi)]_1$,
\item structure de bonne intrication partielle : $\kappa_{BIP}(\psi)=[BIP(\psi)]_1$,
\item structure de mélange total : $\kappa_{MT}(\psi)=[MT(\psi)]_1$,
\item structure d'intrication partielle : $\kappa_{IP}(\psi)=[IP(\psi)]_1$,
\item structure de mélange : $\kappa_{ML}(\psi)=[ML(\psi)]_1$,
\item structure de non claire séparabilité : $\kappa_{NCS}(\psi)=[NCS(\psi)]_1$.
\end{itemize}
\end{df}

Par construction, on a de façon immédiate, pour tout état $\psi\in \bigotimes_{i\in I} H_i$, les relations de finesse suivante entre les structures connectives définies ci-dessus :
\[\kappa_{GI}(\psi)\subset \kappa_{BIP}(\psi)\subset\kappa_{IP}(\psi),\]
\[\kappa_{GI}(\psi)\subset \kappa_{MT}(\psi)\subset\kappa_{IP}(\psi),\]
\[\kappa_{IP}(\psi)\subset \kappa_{ML}(\psi)\subset\kappa_{NCS}(\psi).\]

\subsubsection{Exemples : états $EPR$, $GHZ$ et $O_2$}

\begin{exm}[État $EPR$]\label{exemple etat EPR et desintrication}

Le système quantique considéré est composé de deux particules $\mathcal{S}_1$ et $\mathcal{S}_2$, chacune ayant pour espace d'états un plan $H_i\simeq \mathbf{C}^2$, engendré par des états notés $\vert 0>$ et $\vert 1>$ :
\[H_i=\mathrm{Vect}(\vert 0>,\vert 1>)=\{\alpha \vert 0> +\beta \vert 1>, (\alpha,\beta)\in \mathbf{C}^2\},\] et se trouve dans l'état 
\begin{equation}\label{formule etat EPR}
EPR=\dfrac{\vert 00> + \vert 11>}{\sqrt{2}}.
\end{equation}

Puisque $card(I)=2$, la seule partie de $I$ à considérer est $J=I$. Or, d'après la remarque \ref{rmq I=J}, l'état $EPR$ étant intriqué (non factorisable), $EPR$ est globalement intriqué sur $I$. Par conséquent, toutes les structures de désintrications de $EPR$ coïncident avec la structure connective grossière sur l'ensemble $I$.

\end{exm} 


\begin{exm}[État $GHZ$]\label{exemple etat GHZ et desintrication}

Reprenons l'exemple principal considéré par Aravind \cite{Aravind:1997} : un système de trois qubits est préparé dans l'état  $\psi=GHZ$ défini par la formule (\ref{formule etat GHZ})
\[ GHZ=\dfrac{\vert 000>+\vert 111>}{\sqrt{2}}.
\]
C'est un état intriqué, donc $GHZ$ est globalement intriqué sur $I=\{1,2,3\}$. 

Par contre, $GHZ$ n'est pas globalement intriqué sur $J=\{2,3\}$, puisque une mesure de 
l'observable\footnote{Observable dont les états propres sont précisément ceux notés $\vert 0>$ et $\vert 1>$.} 
$Z=\left(\begin{array}{cc}
1 & 0 \\ 
0 & -1
\end{array} \right)$ sur la première particule transforme de façon équiprobable l'état du système global soit en $\vert 000>$, soit en $\vert 111>$, donc dans tous les cas en des états totalement désintriqués. D'un autre coté, comme le remarque Aravind, une mesure également sur la première particule de 
l'observable  $X=\left(\begin{array}{cc}
0 & 1 \\ 
1 & 0
\end{array} \right),$  dont nous notons $\vec{u}$ et $\vec{v}$ deux vecteurs propres unitaires orthgonaux, conduit à partir de l'état $GHZ$ à deux états possibles équiprobables, $\vec{u}\otimes\dfrac{\vert 00>+\vert 11>}{\sqrt{2}}$ et $\vec{v}\otimes\dfrac{\vert 00>+\vert 11>}{\sqrt{2}}$, d'où $(X\otimes Id_J)_J[GHZ]=\{\dfrac{\vert 00>+\vert 11>}{\sqrt{2}}\}$, qui est intriqué. On en déduit que $GHZ$ est à la fois partiellement bien intriqué et partiellement bien séparable sur $J$. Par symétrie, on aboutit à la même conclusion pour $\{1,3\}$ et $\{1,2\}$, de sorte que les structures connectives de désintrication pour l'état $GHZ$ sont les suivantes :

\begin{itemize}
\item les structures d'intrication globale et de mélange total sont borroméennes,
\item les autres structures de désintrication sont totalement connectées (structures grossières),
\end{itemize}
autrement dit :
\[
\kappa_{GI}(GHZ)=\kappa_{MT}(GHZ)=\mathcal{B}_3,
\]
\[
\kappa_{BIP}(GHZ)=\kappa_{IP}(GHZ)=\kappa_{ML}(GHZ)=\kappa_{NCS}(GHZ)=\mathcal{P}(I).
\]

\end{exm}



\begin{exm}[L'état $O_2$]\label{exm etat 02 ordre deux}
On vérifie  que la structure d'intrication globale $\kappa_{GI}(O_2)$ de l'état 
\begin{equation}\label{formule etat ordre deux}
O_2=\dfrac{2}{\sqrt{13}}(\vert 000> + \vert 011> + \vert 100> -\dfrac{1}{2}\vert 111>)
\end{equation}

est 
\[
\kappa_{GI}(O_2)=\{\emptyset, \{1\},\{2\},\{3\},\{2,3\},\{1,2,3\}\}.
\]

Autrement dit, dans un système de trois particules intriquées selon l'état $O_2$, les particules $2$ et $3$ sont (globalement) intriqués dans le système, et la particule $1$ est globalement intriquée à $\{2,3\}$, mais sans être globalement intriquée spécifiquement ni à $2$ seule, ni à $3$ seule.

Remarquons que cette structure est d'ordre connectif égal à $2$. En section \ref{subsection ordre etat}, cela nous conduira à dire que l'état $O_2$ lui-même est d'ordre connectif $2$.
\end{exm}

\subsection{Structures connectives de densité}\label{subsection structures connectives de densite}

\subsubsection{L'idée de Sugita}

En 2007, Ayumu Sugita \cite{Sugita:2007}, s'appuie sur le formalisme des opérateurs de densité pour montrer que le n\oe ud borroméen exprime intrinsèquement la structure d'intrication de l'état $GHZ$, indépendamment du choix d'une base que constitue le fait de mesurer une observable plutôt qu'une autre, comme l'avait fait Aravind \cite{Aravind:1997}.

Comme nous l'avons rappelé précédemment (section \ref{subsubsec densite composition}), le système composé considéré étant supposé être dans un état quantique donné, le formalisme des opérateurs de densité permet en effet d'attribuer un état de densité à tout sous-système de celui-ci. Pour le dire simplement, il est possible dans ce formalisme d'oublier certains sous-systèmes et de regarder l'état dans lequel se trouve le reste.

Nous nous proposons dans la présente section \ref{subsection structures connectives de densite} de prolonger l'idée de Sugita à tout état quantique de tout système composé d'un nombre fini de sous-systèmes. Comme dans les sections précédentes, on note $I=\{1, 2,...,k\}$ un ensemble fini d'indices, $\mathcal{S}$ est un système quantique d'espace des états $H=\bigotimes_{i\in I} H_i$, composé d'une famille  $(\mathcal{S}_i)_{i\in I=\{1,...,k\}}$ de sous-systèmes ayant chacun pour espaces d'états un espace de Hilbert $H_i$, et l'on supposera que $\mathcal{S}$ se trouve dans un état de densité $\rho\in \mathcal{T}(H)$.

\subsubsection{Corrélation et intrication de $\rho$ sur une partie $J\subset I$}

Dans tout ce qui suit, $J$ désigne une partie de $I$ possédant au moins deux éléments.
Comme précédemment, on note $\neg J$ le complémentaire de $J$ dans $I$ et, pour toute partie $L$ de $I$, on désigne par $H_L$ l'espace de Hilbert
\[
H_L=\bigotimes_{l\in L} H_l.
\]

En nous appuyant sur les définition \ref{df correlation et intrication densite} et \ref{df reduction densite}, nous posons d'abord celle-ci :

\begin{df} Nous dirons que $\rho$ est \emph{complètement corrélé sur $J$}, et l'on notera
\[\rho\in corr(J),\] 
si la réduction $\rho^J$ de $\rho$ à $J$ est complètement corrélée. Nous dirons en outre que $\rho$ est \emph{complètement intriqué sur $J$}, et l'on écrira
\[\rho\in intr(J),\]
si $\rho^J$  est complètement intriqué.
\end{df}

\subsubsection{Structures connectives de densité de $\rho$}

\begin{df}\label{df structures densite} On appelle \emph{structure connective de densité par corrélation} de $\psi$, la structure connective intègre $\kappa_{corr}(\rho)$ engendrée par les parties $J$ de $I$ sur lesquelles $\rho$ est complètement corrélé :
\[
\kappa_{corr}(\rho)=[\{J\in\mathcal{P}I, \rho\in corr(J)\}]_1.
\]
On appelle \emph{structure connective de densité par intrication} de $\psi$, ou plus simplement \emph{structure de Sugita} de $\rho$, la structure connective intègre $\kappa_{S}(\rho)$ engendrée par les parties $J$ de $I$ sur lesquelles $\rho$ est complètement intriqué :
\[
\kappa_S(\rho)=[\{J\in\mathcal{P}I, \rho\in intr(J)\}]_1.
\]
\end{df}

\begin{exm}\label{exm GHZ et densite} On vérifie facilement que l'état $GHZ$ décrit à l'exemple \ref{exemple etat GHZ et desintrication} vérifie
\[
\kappa_S(GHZ)=\mathcal{B}_3.
\]
C'est précisément ce que remarque Sugita dans son article \cite{Sugita:2007}.
\end{exm}

On trouvera d'autres exemples de structures connectives de dispositifs multilocaux dans la présentation placée sur la page web 

\url{https://sites.google.com/site/logiquecategorique/Contenus/CSQE}.

\begin{xrc} Déterminer $\kappa_{corr}(GHZ)$.
\end{xrc}

\subsection{Ordre connectif d'un état quantique intriqué}\label{subsection ordre etat}

\begin{df}\label{df ordre etat} Étant donné $\psi$ un état quantique pur d'un système quantique composé, on appelle \emph{ordre connectif de désintrication de $\psi$}, et l'on note $\Omega_{c}(\psi)$, le maximum des ordres connectifs des diverses structures connectives de désintrication de $\psi$ données dans la définition \ref{df structures de desintrication}. Autrement dit, $\Omega_{c}(\psi)$ est égal à : 
\[\max\{\Omega(\kappa_{GI}(\psi)), \Omega(\kappa_{BIP}(\psi)),\Omega(\kappa_{MT}(\psi)), \Omega(\kappa_{IP}(\psi)),\Omega(\kappa_{ML}(\psi)),\Omega(\kappa_{NCS}(\psi))\}.\]

%
%

Étant donné $\rho$ un état quantique (éventuellement mixte) d'un système quantique composé, état donné par un opérateur de densité $\rho$, on appelle \emph{ordre connectif de densité de $\rho$}, et l'on note $\Omega_{f}(\psi)$, le maximum des ordres connectifs des deux structures connectives de densité de $\rho$ données dans la définition \ref{df structures densite} : 

\[\Omega_f(\rho)=\max\{\Omega(\kappa_{corr}(\rho)), \Omega(\kappa_{S}(\rho))\}.\]

Enfin, on appelle \emph{ordre connectif d'un état pur $\psi$}, et l'on note $\Omega(\psi)$, le maximum de son ordre connectif de désintrication  et de son ordre connectif de densité :
\begin{equation}\label{formule ordre connectif etat pur}
\Omega(\psi)=\max(\Omega_c(\psi),\Omega_f(\vert\psi><\psi\vert)).
\end{equation}
\end{df}

\begin{exm}\label{exm etat ordre 2} On vérifie sans peine que les états $EPR$ et $GHZ$ sont d'ordre connectif égal à $1$. Pour trois particules intriquées, l'ordre connectif est néces\-sairement inférieur ou égal à $2$. Un exemple d'état quantique intriquant trois particules et d'ordre connectif $2$ est donné par l'état $O_2$ défini par la formule (\ref{formule etat ordre deux}) de l'exemple \ref{exm etat 02 ordre deux}.
\end{exm}

\subsection{Conjectures brunnienne}

Nous conjecturons que toutes les structures connectives définies précédemment satisfont la conjecture brunnienne correspondante, à savoir, comme nous l'avons indiqué dans la section \ref{subsection conjectures brunniennes} page \pageref{subsection conjectures brunniennes} :

\emph{Pour toute structure connective finie $\kappa$, il existe un système quantique $\mathcal{S}$ composé et un état quantique $e$ de ce système, tel que la structure connective de $e$ soit précisément $\kappa$.}

\section{Structures relationnelles des dispositifs multi-locaux}\label{section structure relationnelle des dispositifs}\sectionmark{Dispositifs multilocaux}

Reprenant, avec quelques changement terminologiques, le point de vue de Christophe Chalons dans son travail sur les degrés ludiques (voir \cite{Chalons:201405}), nous formalisons par la notion de \emph{dispositif multilocal} les expériences de mesure portant sur un système quantique intriqué, de telles expériences étant identifiées à des familles de questions \emph{locales} posées au système par des expérimentateurs distants les uns des autres. 

Regroupées sous l'appellation générale de \emph{structures connectives relationnelles}, nous définissons ensuite plusieurs types de structures connectives pour les dispositifs multilocaux, à savoir diverses structures qualifiées de \emph{tensorielles}, et plusieurs structures dites \emph{domaniales}. Renvoyant à un travail ultérieur, nous signalons également la notion de structure connective ludique. 

\`{A} la fin de la présente section \ref{section structure relationnelle des dispositifs}, nous introduisons la notion d'ordre connectif d'un dispositif multi-local.

\subsection{Définition des dispositifs multilocaux}

\subsubsection{Relations binaires vues comme applications multivalentes}

Étant donnée une relation binaire $R$ d'un ensemble $A$ vers un ensemble $B$, nous noterons $(a,b)\in R$ pour exprimer que $a$ est en relation avec $b$, notation qui revient à identifier $R$ avec son graphe.

En outre, nous identifierons couramment $R$ à l'application $A\rightarrow \mathcal{P}B$ qui à tout $a\in A$ associe l'ensemble $R(a)$ --- éventuellement vide --- de tous les $b\in B$ tels que $(a,b)\in R$ :
\[
R(a)=\{b\in B, (a,b)\in R\}.
\]

Pour souligner ce point de vue multivalent sur les relations binaires, nous écrirons 
\begin{equation}\label{formule relation binaire comme transition}
R:A\rightsquigarrow B
\end{equation} pour exprimer que $R$ est une telle relation binaire.

\subsubsection{Dispositifs expéri\-mentaux (globaux)} 

\begin{df}[Dispositif expéri\-mental (global)]\label{df dispositif experimental} Nous appellerons \emph{dispositif expéri\-mental global}, ou simplement \emph{dispositif global}, ou encore plus simplement \emph{dispositif}, toute relation binaire $D:Q\rightsquigarrow R$, l'ensemble $Q$, supposé non vide, étant appelé l'ensemble des \emph{expé\-riences} ou des \emph{questions} du dispositif, tandis que $R$ est l'ensemble des \emph{résultats} ou des \emph{réponses} du dispositif. Nous dirons qu'un dispositif $D$ est co\-hé\-rent si pour tout $q\in Q$, on a $D(q)\neq\emptyset$.
\end{df}

Dans la suite, tous les dispositifs considérés seront implicitement (et parfois explicitement) supposés co\-hé\-rents. 


\begin{df}[Descriptions et réalisations d'un dispositif expéri\-mental]\label{df descriptions des dispositifs} Étant donnée $D:Q\rightsquigarrow R$ un dispositif expéri\-mental, on appelle \emph{garantie} ou \emph{description} de $D$ toute relation binaire $G:Q\rightsquigarrow R$ telle que $D\subset G$, au sens où pour tout $q\in Q$, on ait $D(q)\subset G(q)$. Dans ce cas, nous dirons aussi que $D$ est une \emph{réalisation partielle} de $G$. 
\end{df}
 
\begin{rmq} Puisque nous utilisons le symbole $\subset$ au sens large --- il a donc pour nous exactement la même signification que le symbole $\subseteq$ ---  il est clair que, les ensembles $Q$ et $R$ étant donnés, la relation $D\subset G$ est une relation d'ordre sur l'ensemble des dispositifs $Q\rightsquigarrow R$. En particulier, cette relation est réflexive, de sorte que   
$D$ se décrit lui-même, constituant ce que nous appellerons la \emph{description exacte} de $D$. 
\end{rmq} 

Intuitivement, la différence entre une description quelconque d'un dispositif expéri\-mental et la description exacte de ce dispositif, est qu'une description quelconque indique des possibilités qui pourraient ne jamais se réaliser, tandis que toutes les possibilités données par la description exacte doivent effectivement pouvoir se réaliser.

Étant donnée $\mathcal{D}$ un ensemble de dispositifs expéri\-mentaux portant sur les mêmes expé\-riences $Q$ et à valeur dans le même ensemble de résultats $R$, nous noterons $\bigcup_{D\in \mathcal{D}}D$,
 ou plus simplement $\bigcup_{\mathcal{D}}$, la borne supérieure de la famille $\mathcal{D}$ pour la relation d'ordre $\subset$, autrement dit le dispositif expéri\-mental $S:Q\rightsquigarrow R$ défini pour tout $q\in Q$ par $S(q)=\bigcup_{D\in \mathcal{D}}D(q)$.

\begin{df} On dit qu'un dispositif expéri\-mental $f: Q\rightsquigarrow R$ est \emph{déterministe} si pour tout $q\in Q$ l'ensemble $f(q)$ est un singleton. Dans ce cas, on identifiera $f$ avec l'application $f:Q\rightarrow R$ qui à tout $q\in Q$ associe l'unique élément $r$ de $f(q)$, et nous noterons souvent, s'il n'y a pas de risque de confusion, $f(q)=r$. 
\end{df}

Conformément aux notations précédentes, si $\mathcal{F}$ est un ensemble de dispositifs expéri\-mentaux \emph{déterministes} portant sur les mêmes expé\-riences $Q$ et à valeur dans le même ensemble de résultats $R$, $\bigcup_{\mathcal{F}}$ désignera le dispositif expéri\-mental $D$ défini pour toute expé\-rience $q\in Q$ par $D(q)=\{r\in R,\exists f\in\mathcal{F}, f(q)=r\}$.
 
 \begin{prop}\label{prop realisations partielles deterministes} Si on admet l'axiome du choix (ou si l'ensemble des questions $Q$ est fini), alors pour tout dispositif expéri\-mental co\-hé\-rent $D:Q\rightsquigarrow R$, on a $\bigcup_\mathcal{F}=D$, où $\mathcal{F}$ désigne l'ensemble des réalisations partielles déterministes de $D$.
 \end{prop}
 
\subsubsection{Dispositifs multilocaux}

\begin{df} Etant donné $k$ un entier, 
on appelle \emph{dispositif expéri\-mental multilocal d'uplicité $k$}, ou simplement \emph{dispositif}, tout dispositif expéri\-mental $D:Q\rightsquigarrow R$ avec $Q$ de la forme 
$Q=Q_1\times\cdots\times Q_k$ et $R$ de la forme $R=R_1\times \cdots R_k$, où  $Q_1$, $\cdots$, $Q_k$ sont des ensembles non vides, de même que  $R_1$, $\cdots$ , $R_k$.
\end{df}

Posant $I=\{1,\cdots, k\}$, nous noterons $q_i$ la $i$\up{ème} composante de tout $q\in Q=\prod_{i\in I} Q_i$, de sorte que $q=(q_1,\cdots,q_k)$. Plus généralement, pour $J\subset I$, nous noterons $q_{\vert J}$, ou simplement $q_J$, la famille extraite de $q$ en ne retenant que les composantes d'indices $j\in J$ :
\[
q_{\vert J}=(q_j)_{j\in J}.
\]

\begin{df}[Sous-dispositifs] Soit $D:\prod_{i\in I} Q_i=Q\rightsquigarrow R=\prod_{i\in I} R_i$ un dispositif d'uplicité $k$, où $I=\{1,\cdots,k\}$, et soit $J\subset I$ un ensemble non vide d'indices. On appelle \emph{sous-dispositif} de $D$ pour $J$ le dispositif $\prod_{j\in J} Q_j\rightsquigarrow \prod_{j\in J} R_j$ noté $D_{[J]}$  et défini pour tout $(q_j)_{j\in J}$ par 
\[
D_{[ J]}((q_j)_{j\in J})=
\{(r_j)_{j\in J}\in \prod_{j\in J} R_j, 
\exists \tilde{q}\in Q, 
\exists \tilde{r}\in D(\tilde{q}),
(q_j)_{j\in J}=\tilde{q}_{\vert J}\,\mathrm{et}\,(r_j)_{j\in J}=\tilde{r}_{\vert J}\}
\]
En particulier, lorsque $J$ est un singleton $J=\{j\}$, $D_{[j]}$ est le sous-dispositif local de $D$ en $j$.
\end{df}

\begin{rmq} En général, un sous-dispositif d'un dispositif déterministe n'est pas lui-même déterministe\footnote{C'est bien ce qui légitime de partir à la recherche de variables cachées lorsqu'on cherche à rendre compte de façon déterministe d'un processus qui n'apparaît pas tel.}. On fera par ailleurs attention au fait que, pour tout dispositif déterministe $f:\prod_{i\in I} Q_i=Q\to R=\prod_{i\in I} R_i$ et tout indice $i$, la $i$\up{ème} composante $f_i$ de $f$ est une application $f_i:Q\to R_i$ qu'il s'agit de ne pas confondre avec le dispositif local $f_{[i]}:Q_i\rightsquigarrow R_i$. On a en effet
\[
f_i(q)=f(q)_{\vert i},
\]
tandis que
\[
f_{[i]}(q_i)=\{r_i\in R_i, \exists \tilde{q}\in Q, q_i=\tilde{q}_{\vert i} \,\mathrm{et}\, 
r_i=f(\tilde{q})_{\vert i}\}.
\]
\end{rmq}

La seule connaissance des dispositifs locaux ne permet pas, en général, de retrouver le dispositif multilocal d'où ils proviennent. Lorsque néanmoins cela se produit, nous dirons que le dispositif multilocal en question est \emph{complètement local}, ou encore, si $k\geqslant 2$, \emph{complètement séparable}. Pour préciser, à la section suivante, ce type de notions, nous aurons d'abord besoin de définir le produit tensoriel\footnote{L'expression \emph{tensoriel} fait ici  référence notamment au fait que ce produit se traduit par l'augmentation du nombre des variables.} de plusieurs dispositifs multilocaux :

\begin{df} Étant donnée une famille $(D_i:Q_i\rightsquigarrow R_i)_{i\in I}$ de dispositifs co\-hé\-rents, où $I=\{1,\cdots,k\}$, on appelle \emph{produit tensoriel} de cette famille le dispositif multi-local $D=\bigotimes_{i\in I} D_i:\prod_{i\in I}Q_i\rightsquigarrow \prod_{i\in I} R_i$ défini par $D((q_i)_{i\in I})=\prod_{i\in I} D_i(q_i)$, où l'expression $\prod_{i\in I} D_i(q_i)$ désigne le produit cartésien des ensembles $D_i(q_i)$.
\end{df}

Bien entendu, si $D=\bigotimes_{i\in I} D_i$, les sous-dispositifs locaux $D_{[i]}$ de $D$ coïncident avec les $D_i$.

\subsection{Expériences quantiques en termes de dispositifs multilocaux}\label{section experiences quantiques comme dispositifs}

La structure relationnelle des dispositifs de mesure de systèmes quantiques intriqués peut être décrite, au moins  partiellement, en termes de dispositifs multilocaux : un système quantique constitué de $k$ sous-systèmes ayant été préparé dans un état donné d'intrication, on imagine que chacun de ces sous-systèmes $\mathcal{S}_i$ est expédié vers un expérimentateur qui se trouve en un lieu $i$ distant des autres et qui peut réaliser sur $\mathcal{S}_i$ une expérience $q_i$ appartenant à un ensemble donné d'expériences $Q_i$, obtenant une réponse $r_i$ appartenant à un ensemble $R_i$ de réponses possibles. Le dispositif multilocal $D$ décrivant cette expérience est alors, pour chaque famille $(q_1,\cdots, q_i,\cdots, q_k)$ d'expériences locales, l'ensemble $D((q_i)_{i\in I})$ de toutes les familles $(r_1,\cdots, r_k)$ de réponses possibles.

Bien entendu, le dispositif multilocal $D$ ne décrit qu'un aspect du système intriqué $\mathcal{S}$, à savoir la manière dont celui-ci répond, dans des conditions données, à des expériences locales, alors même que l'on a fait l'hypothèse simplificatrice d'une non-évolution continue des états du systèmes. 

\begin{rmq}[Respect de la contrainte relativiste]\label{rmq respect contrainte relativiste}
Soulignons que, d'après Chalons \cite{Chalons:201405}, les dispositifs multilocaux définis par des expériences de mesure portant sur des systèmes quantiques intriqués sont toujours compatibles avec la contrainte relativiste sur la transmission d'information  : ils ne permettent en aucun cas de transmettre une information d'un expérimentateur à un autre. Par contre, une réalisation déterministe $f$ d'un tel dispositif peut ne pas respecter la contrainte relativiste. Tout se passe donc comme si, à chaque fois que l'on fait l'expérience décrite par un tel dispositif, une réalisation déterministe était choisie aléatoirement par la nature, sans que les expérimentateurs puissent jamais savoir laquelle.
\end{rmq}


\begin{exm}[$D_{EPR}$]\label{exm dispositif DEPR} On considère le système quantique à deux particules intriquées décrit à l'exemple \ref{exemple etat EPR et desintrication} page \pageref{exemple etat EPR et desintrication}.

Dans la suite, nous utiliserons les notations alternatives suivantes :

\[
\vec{x}=\vert 0>
\] et 
\[
\vec{y}=\vert 1>,
\] 
en distinguant en outre par un indice le vecteur $\vec{x}_1\in H_1$ du vecteur $\vec{x}_2\in H_2$, et de même pour $\vec{y}$.

Conformément à la formule (\ref{formule etat EPR}), le système se trouve dans l'état $\psi=EPR$ défini par
\[
\psi=\dfrac{\vert 00>+\vert 11>}{\sqrt{2}}=\dfrac{\vec{x}_1\otimes \vec{x}_2+\vec{y}_1\otimes \vec{y}_2}{\sqrt{2}}=
\dfrac{\vec{x}_1 \vec{x}_2+\vec{y}_1 \vec{y}_2}{\sqrt{2}},
\]

Le dispositif que nous noterons $D_{EPR}$ correspond alors à l'expérience suivante : une fois éloignées l'une de l'autre, on mesure pour chacune des deux particules une observable représentée par la \og matrice de Pauli\fg\, traditionnellement notée $Z$ (parce qu'elle correspond à une mesure de spin dans la direction de l'axe des $z$), à savoir
$Z=\left(\begin{array}{cc}
1 & 0 \\ 
0 & -1
\end{array} \right)$. On pourra éventuellement préciser  $Z_1$ ou $Z_2$ selon que l'observable $Z$ se rapporte au sous-système $\mathcal{S}_1$ ou au sous-système $\mathcal{S}_2$.

Pour fixer les idées, commençons
\footnote{\label{note 1 chronologie fictive} Du fait de la relativité, cela n'a en fait pas de sens de dire que l'une des deux mesures serait faite avant l'autre (voir à ce sujet la remarque \ref{rmq chronologie fictive des mesures et des desintrication}).}
 par examiner l'effet de cette mesure (partielle) sur la particule $\mathcal{S}_1$. En tant que mesure globale, il s'agit de celle définie par l'opérateur $Z_1\otimes Id_2$, où $Id_2$ désigne l'identité de $H_2$. Les sous-espaces propres de $Z_1\otimes Id_2$ sont $\vec{x}_1\otimes H_2$ et $\vec{y}_1\otimes H_2$, de valeurs propres respectives $1$ et $-1$. Lors de la mesure correspondante, le vecteur $\psi$ est projeté 
\begin{itemize}
\item soit, avec une probabilité $\dfrac{1}{2}$, sur $\dfrac{\vert 00>}{\sqrt{2}}$ --- donc après renormalisation sur l'état 
${\vert 00>}$ --- 
en donnant comme résultat de mesure la valeur propre $1$,
\item soit, également avec une probabilité $\dfrac{1}{2}$, sur $\dfrac{\vert 11>}{\sqrt{2}}$ --- 
donc après renormalisation sur l'état ${\vert 11>}$ --- 
 avec comme résultat la valeur propre $-1$.
\end{itemize}

Considérons ensuite\footnote{\label{note 2 chronologie fictive} Mêmes remarques que dans la note \ref{note 1 chronologie fictive}.} l'effet d'une mesure de l'observable $Z$ sur la deuxième particule, autrement dit d'une mesure de l'opérateur  $Id_1\otimes Z_2$.
Les sous-espaces propres de cet opérateur étant $H_1\otimes \vec{x}_2$ et $H_1\otimes \vec{y}_2$ avec comme valeurs propres respectives $1$ et $-1$, on a finalement le résultat suivant :
\begin{itemize}
\item si cette seconde mesure est effectuée sur l'état global ${\vert 00>}$, celui-ci reste inchangé avec une probabilité égale à $1$, produisant comme résultat la valeur propre $1$,
\item si elle est effectuée sur l'état global ${\vert 11>}$, celui-ci reste également inchangé avec une probabilité égale à $1$, produisant comme résultat la valeur propre $-1$.
\end{itemize}

\noindent\textbf{Récapitulatif}. Finalement, en réponse à l'unique question posée par chacun des deux expérimentateurs, le dispositif $D_{EPR}$ donne pour réponse soit $(1,1)$, soit $(-1,-1)$.
 Autrement dit, \emph{avec un codage légèrement différent}, nous avons obtenu le dispositif multilocal $D_{EPR}$ défini par 
\[Q_1=Q_2=\{*\},\] 
\[R_1=R_2=\{0,1\}\] et \[D_{EPR}(**)=\{00,11\}.\]

\begin{rmq}[Sur l'aspect fictif de la chronologie des mesures]\label{rmq chronologie fictive des mesures et des desintrication} Comme indiqué dans les notes \ref{note 1 chronologie fictive} et \ref{note 2 chronologie fictive}, le fait de \emph{commencer} par examiner l'impact d'une mesure sur la particule $1$ \emph{puis} de considérer celui d'une autre mesure sur la particule $2$ ne saurait traduire une chronologie réelle dont la Relativité conteste d'ailleurs l'existence. La même remarque vaut pour toute description de l'effet d'une mesure quantique qui ferait, ne serait-ce qu'implicitement, appel à une notion de simultanéïté incompatible avec la relativité. Certes, les mesures effectuées peuvent conduire à une désintrication du système, mais il n'est pas possible de dire, lorsque plusieurs mesures distantes sont effectuées sur les parties d'un système intriqué, que l'une de ces mesures serait davantage qu'une autre la cause de cette désintrication. On peut néanmoins vérifier que l'ordre fictif que nous utilisons pour décrire l'ensemble du processus n'a pas d'incidence sur le résultat obtenu, de sorte que nous pouvons toujours introduire arbitrairement un tel ordre. En effet, la projection orthogonale successive d'un vecteur d'état sur des sous-espaces propres de l'espace $H$ associés à une suite d'observables partielles $A_1$, $A_2$, $A_3$,..., sous-espaces de la forme $E_{\lambda_1}\otimes H_2\otimes H_3...$, puis $H_1\otimes E_{\lambda_2}\otimes H_3...$, puis $H_1\otimes H_2\otimes E_{\lambda_3}...$, etc., revient finalement à projeter orthogonalement sur les sous-espaces de la forme $E_{\lambda_1}\otimes E_{\lambda_2}\otimes E_{\lambda_3}...$, à savoir les sous-espaces propres  d'une unique observable $A=\bigotimes_{i\in I} A'_i$ rassemblant toutes les observables partielles $A_i$ considérées, quitte à avoir remplacé chaque opérateur $A_i$ par un opérateur $A'_i$ de mêmes sous-espaces propres mais dont les valeurs propres soient telles qu'on puisse les retrouver arithmétiquement à partir de celles de $A$.
\end{rmq}

\end{exm}


\begin{exm}[$D_{EPR,2}$ Dispositif $EPR$ à deux questions]\label{exm DEPR2 dispositif} Le système quantique considéré est le même que dans l'exemple \ref{exm dispositif DEPR} ci-dessus, mais cette fois chaque expérimentateur aura le choix entre \emph{deux} mesures différentes : outre celle de l'observable $Z=\left(\begin{array}{cc}
1 & 0 \\ 
0 & -1
\end{array} \right)$, il pourra également choisir de mesurer l'observable définie par la \og matrice de Pauli\fg traditionnellement notée $X$, à savoir :
$X=\left(\begin{array}{cc}
0 & 1 \\ 
1 & 0
\end{array} \right).$

Aux notations de l'exemple \ref{exm dispositif DEPR}, nous ajoutons les suivantes pour désigner les vecteurs propres de $X$ de valeurs propres respectives $1$ et $-1$ :
\[
\vec{u}=\dfrac{\vec{x}+\vec{y}}{\sqrt{2}}
\] et 
\[
\vec{v}=\dfrac{\vec{x}-\vec{y}}{\sqrt{2}},
\]
de sorte que l'on a $\vec{x}=\dfrac{\vec{u}+\vec{v}}{\sqrt{2}}$ et $\vec{y}=\dfrac{\vec{u}-\vec{v}}{\sqrt{2}}$.

En examinant l'effet d'une mesure de $Z$ sur le sous-système $1$ \og suivi\fg\footnote{Voir la remarque \ref{rmq chronologie fictive des mesures et des desintrication}.} soit d'une mesure de $Z$ sur le sous-système $2$ soit d'une mesure de $X$ sur le sous-système $2$, et en remarquant que les deux autres cas de figure se ramènent à celui-ci par symétrie, on obtient une description de cette expérience de mesure en  termes d'un dispositif multilocal que nous noterons $EPR_2$, à savoir, après un changement de code évident\footnote{\label{note codage resultats} Qui consiste d'une part à coder le fait que l'on mesure l'observable $Z$ par un $0$, et l'observable $X$ par un $1$, et d'autre part, en ce qui concerne les valeurs obtenues pour ces mesures, à remplacer la valeur propre $-1$ par le code $0$.} :

\mbox{}

\[Q_1=Q_2=\{0,1\},\] \[R_1=R_2=\{0,1\},\] 

\[D_{EPR,2}(00)=D_{EPR,2}(11)=\{00,11\}\]
et
\[D_{EPR,2}(01)=D_{EPR,2}(10)=R_1\times R_2.\]

\begin{rmq}\label{rmq matrices de Pauli modifiee avec vp 0 et 1}
Au lieu d'effectuer, pour des raisons pratiques de lisibilité, le changement de code précisé dans la note \ref{note codage resultats}, on aurait pu d'emblée utiliser, au lieu des matrices de Pauli  $Z$ et $X$, les matrices suivantes, de mêmes sous-espaces propres : $Z'=\left(\begin{array}{cc}
0 & 0 \\ 
0 & 1
\end{array} \right)$ et 
$X'=\dfrac{1}{2}\left(\begin{array}{cc}
1 & -1 \\ 
-1 & 1
\end{array} \right).$ Apparemment, bien que rien ne l'interdise dans le formalisme, l'habitude n'est pas d'utiliser, dans la définition des observables donnant lieu à mesures projectives, des matrices non régulières, et pour le moment nous nous en tiendrons à cette habitude.
\end{rmq}
 
\end{exm}


\begin{exm}[$D_{GHZ}$, un dispositif pour l'état $GHZ$]\label{exm dispositif GHZ}

On considère le système quantique de l'exemple \ref{exemple etat GHZ et desintrication} page \pageref{exemple etat GHZ et desintrication}, constitué de trois qubits intriqués dans l'état $GHZ$ défini par la formule \ref{formule etat GHZ},
et, comme pour l'exemple \ref{exm DEPR2 dispositif}, on suppose que chaque expérimentateur a le choix entre deux requêtes locales, à savoir la mesure soit de l'observable $Z=\left(\begin{array}{cc}
1 & 0 \\ 
0 & -1
\end{array} \right)$, soit  de l'observable 
$X=\left(\begin{array}{cc}
0 & 1 \\ 
1 & 0
\end{array} \right),$ dont les vecteurs propres sont notés comme précédemment (voir la figure \ref{figure GHZD} page \pageref{figure GHZD}).

On vérifie facilement que le choix de mesurer les trois observables $(Z_1,Z_2,Z_3)$ conduit, avec équiprobabilité :
\begin{itemize}
\item soit à l'état $\vert 000>$, les résultats de mesure étant $(+1,+1,+1)$,
\item soit à l'état $\vert 111>$, les résultats de mesure étant $(-1,-1,-1)$,
\end{itemize}
résultats que, conformément à la remarque \ref{rmq matrices de Pauli modifiee avec vp 0 et 1}, nous coderons respectivement par $111$ et $000$.

De même, le choix de mesurer soit les trois observables $(Z_1,Z_2,X_3)$, soit $(Z_1,X_2,Z_3)$, soit $(X_1,Z_2,Z_3)$ conduit à un ensemble de quatre résultats possibles, à savoir, dans le cas de $(Z_1,Z_2,X_3)$ et toujours avec le même codage (la valeur propre $-1$ est codée par un $0$) :
\[\{000,001,110,111\},\] les résultats pour les deux autres jeux d'expériences étant obtenus par rotations.

On vérifie enfin que le choix de mesurer trois observables contenant au moins deux fois celle notée $X$ conduit, de façon équiprobable, à l'ensemble des huit résultats possibles $\{0,1\}^3$.

\begin{figure} 
\begin{center}
\includegraphics[bb=0 0 1420 2052,width=12cm]{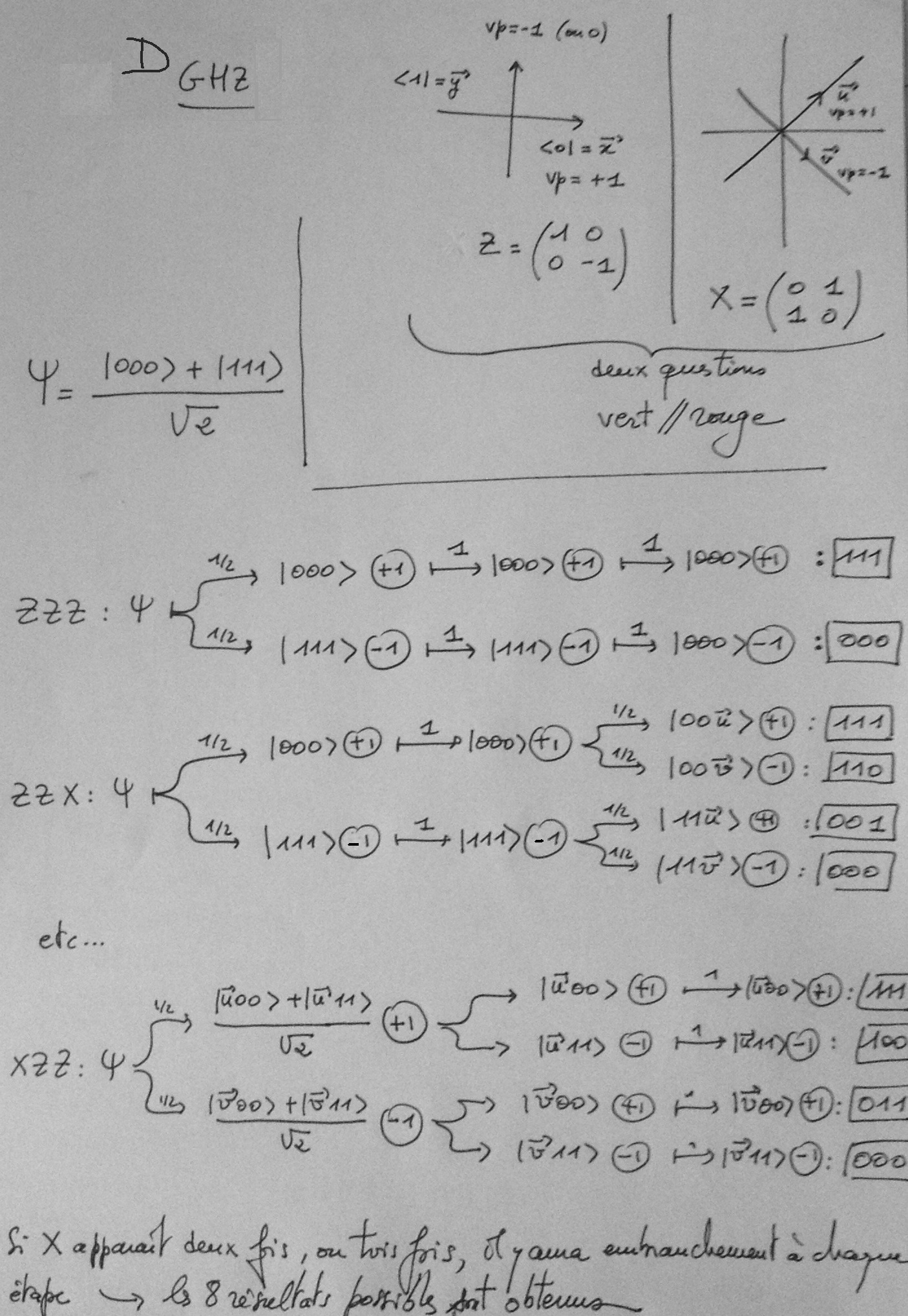}
\caption{Fonctionnement du dispositif $D_{GHZ}$}
\label{figure GHZD}
\end{center}
\end{figure}

\begin{rmq}
Attention de ne pas confondre le \emph{résultat} des mesures, donné par les valeurs propres des observables, avec l'\emph{état quantique final} dans lequel se retrouve le système intriqué au terme de ces mesures. Par exemple, pour une mesure de $(Z_1,Z_2,X_3)$, le résultat est $111$ si le système se retrouve dans l'état $\vert 00u>=\vec{x}_1\otimes \vec{x}_2\otimes \vec{u_3}$. 
\end{rmq}

\end{exm}

\begin{exm}[Dispositif $D_K$ pour l'état $K$]\label{exm dispositif DK}

On appelle $D_K$ le dispositif consistant à faire les \emph{mêmes} expériences de mesure que celles de l'exemple \ref{exm dispositif GHZ}, le système étant cette fois préparé dans l'état quantique $K$ suivant :
\begin{equation}\label{formule etat K}
K=\dfrac{\vert 111>-\vert 001>-\vert 010>-\vert 100>}{2},
\end{equation}

On vérifie sans difficulté que le dispositif $D_K$ ainsi défini répond, à des variantes de codage près, à la description  donnée par  Christophe Chalons (\cite{Chalons:201405}, chapitre 5) de ce qu'il appelle \emph{téléphone $GHZ$}\footnote{Le dispositif en question, malgré cette appellation, ne semble pas entretenir de relation particulière avec l'état $GHZ$ défini ici par la formule \ref{formule etat GHZ}.} à savoir :



\[Q_1=Q_2=Q_3=\{0,1\},\] 
\[R_1=R_2=R_3=\{0,1\},\]
\[D_K(111)=\{001,010,100,111\},\] \[D_K(001)=D_K(010)=D_K(100)=\{000,011,101,110\}\] et, pour les autres expériences $(abc)\in \prod_i Q_i$ : \[D_K(abc)=\prod R_i.\]

%

\end{exm}

\subsection{Notions de localité (ou de séparabilité) pour les dispositifs}

\subsubsection{Définitions}

Le cas de l'uplicité $k=1$ est laissé de coté dans les définitions qui suivent, puisque dans ce cas particulier les dispositifs pourraient être dits à la fois locaux, globaux et non séparables.

\begin{df}[Notions de localité]\label{df localites dispositifs} Soit  $G:\prod_{i\in I} Q_i=Q\rightsquigarrow R=\prod_{i\in I} R_i$ un  dispositif multilocal co\-hé\-rent, avec $I=\{1,\cdots,k\}$, où $k$ est un entier donné supérieur ou égal à $2$.
\begin{itemize}
\item On dit que $G$ est \emph{local} ou \emph{complètement local} ou \emph{complètement séparable} s'il existe une famille de dispositifs (né\-ces\-saire\-ment co\-hé\-rents) $(L_i:Q_i\rightsquigarrow R_i)_{i\in I}$ dont $G$ soit le produit tensoriel : 
\[G=L_1\otimes L_2\otimes \cdots\otimes L_k,\] autrement dit si l'on a 
\[G=G_{[1]}\otimes G_{[2]}\otimes\cdots\otimes G_{[k]},\]
\item On dit que $G$ est \emph{quasi-local} s'il peut s'écrire comme la borne supérieure, autrement dit comme l'union, d'une famille de réalisations partielles co\-hé\-rentes complè\-tement locales, ou, de façon équivalente\footnote{Du fait la proposition \ref{prop realisations partielles deterministes}.}, s'il peut s'écrire comme l'union d'une famille de réalisations partielles déterministes complè\-tement locales,
\item On dit que $G$ est \emph{partiellement local} s'il admet une réalisation partielle co\-hé\-rente $D$ qui soit complè\-tement locale, ou, de façon équivalente\footnote{Du fait la proposition \ref{prop realisations partielles deterministes}.}, s'il admet une réalisation partielle déterministe complètement locale.
\end{itemize}
\end{df}

Peut-être plus utiles sont les définitions suivantes de semi-localité, que nous exprimerons également en terme de séparabilité.

\begin{df} $G$ désignant toujours un dispositif multilocal co\-hé\-rent  $\prod_{i\in I} Q_i=Q\rightsquigarrow R=\prod_{i\in I} R_i$ avec $I=\{1,\cdots,k\}$ et $k\geqslant 2$,
\begin{itemize}
\item On dit que $G$ est \emph{séparable} ou \emph{semi-local} s'il existe une partition
\footnote{Voir la note \ref{note partition}.}
 $J_1\cup J_2$ de $I$ telle que $G$ s'exprime, à l'ordre des facteurs près, comme le produit tensoriel $G=L_{1}\otimes L_{2}$ de deux dispositifs $L_1:\prod_{i\in J_1} Q_i\rightsquigarrow \prod_{i\in J_1} R_i$ et $L_2:\prod_{i\in J_2} Q_i\rightsquigarrow \prod_{i\in J_2} R_i$, autrement dit si l'on a, à l'ordre des facteurs près :
\[G=G_{[J_1]}\otimes G_{[J_2]}.\]
 
 \item  On dit que $G$ est \emph{quasi-séparable} ou \emph{quasi-semi-local} s'il existe une partition $J_1\cup J_2$ de $I$ telle que $G$ soit l'union d'une famille de réalisations partielles $L$ séparables selon $J_1$ et $J_2$, autrement dit vérifiant toutes $L=L_{[J_1]}\otimes L_{[J_2]}$,
 
\item On dit que $G$ est \emph{pseudo-séparable} ou \emph{pseudo-semi-local} s'il peut s'écrire comme l'union d'une famille de réalisations partielles co\-hé\-rentes séparables (mais non né\-ces\-saire\-ment séparables selon la \emph{même} partition),
 
\item On dit que $G$ est \emph{partiellement séparable} ou \emph{partiellement semi-local} s'il admet une réalisation partielle co\-hé\-rente $D:\prod_{i=1}^{i=k} Q_i\rightsquigarrow \prod_{i=1}^{i=k} R_i$  qui soit séparable, ou, de façon équivalente, s'il admet une réalisation déterministe séparable.

\end{itemize}
\end{df}

\pagebreak[3]
\begin{prop} \label{prop implications entre localites} On a les implications suivantes :

\mbox{}

\begin{tabular}{ccccc}
 local &  $\Rightarrow$ & quasi-local &  $\Rightarrow$ & partiellement local \\ 
 $\Downarrow$ &   &  $\Downarrow$ &   &  $\Downarrow$ \\ 
séparable &  $\Rightarrow$ & quasi-séparable &  $\Rightarrow$ & partiellement séparable \\ 
\end{tabular} 

\mbox{}

De plus, quasi-séparable $\Rightarrow$ pseudo-séparable $\Rightarrow$ partiellement séparable.
\end{prop}

\begin{prop} Les dispositifs locaux associés à un dispositif multi-local le caractérisent si et seulement si celui-ci est local.
\end{prop}

\subsubsection{Cas des dispositifs multilocaux déterministes}

On a déjà indiqué, au sein même de la définition \ref{df localites dispositifs}, le fait que la proposition \ref{prop realisations partielles deterministes} entraine la suivante :

\begin{prop} Un dispositif multilocal est partiellement local (resp. partiellement séparable) si et seulement s'il possède une réalisation \emph{déterministe} locale (resp. partiellement séparable). 
\end{prop}

Examinons donc de plus près les dispositifs déterministes du point de vue de la localité.
On sait qu'un dispositif multilocal déterministe $f:\prod_{i\in I} Q_i=Q\to R=\prod_{i\in I} R_i$ peut toujours s'écrire sous la forme $f=(f_1,\cdots,f_k)$, chacune des fonctions $f_i$ étant \emph{a priori} définie sur $Q=\prod_{i\in I} Q_i$, donc étant une fonction de $k$ variables. La définition suivante permet de préciser les variables dont ces fonctions dépendent effectivement.

\begin{df}[Domaine indicial de dépendance]\label{df domaine indicial de dependance} Étant donné un entier $j\in I=\{1,...,k\}$, on dit qu'une application $g:\prod_{i\in I} Q_i=Q\to S$  \emph{ne dépend pas de $j$} si pour tout $\hat{q}=(q_1,q_2,...,q_{j-1},q_{j+1},...,q_k)\in \prod_{i\neq j} Q_i$, l'application 
\[
g_{\hat{q}}:Q_j\ni q_j\mapsto g_{\hat{q}}(q_j)=g(q_1,...,q_j,...,q_k)\in S
\] est constante. Dans le cas contraire --- autrement dit s'il existe 
\[
\hat{q}={(q_1,q_2,...,q_{j-1},q_{j+1},...,q_k)}\in \prod_{i\neq j} Q_i
\]
 et deux éléments $q_j$ et $q'_j$ de $Q_j$ tels que $g(q_1,...,q_j,...,q_k)\neq g(q_1,...,q'_j,...,q_k)$ --- nous dirons que $g$ dépend de $j$. On appelle \emph{domaine indicial de dépendance} de $g$, et l'on note $\mathrm{do}(g)$, l'ensemble des indices $j$ dont $g$ dépend.
\end{df}

\begin{rmq} Un dispositif multilocal déterministe est constant si et seulement si son domaine indicial de dépendance est vide.
\end{rmq}

On a alors les propriétés suivantes :

\begin{prop} Pour un dispositif multilocal déterministe $f:\prod_{i\in I} Q_i=Q\to R=\prod_{i\in I} R_i$, il n'y a pas de différence entre le fait d'être local, quasi-local ou partiellement local, et cela se produit si et seulement si pour tout $i\in I$, $\mathrm{do}(f_i)\subset\{i\}$, autrement dit si et seulement s'il existe une famille d'applications $(g_i:Q_i\to R_i)_{i\in I}$ telle que pour tout $(q_i)_{i\in I}\in Q$ on ait 
\[f((q_i)_{i\in I})=(g_i(q_i))_{i\in I}.\] 
De même, pour un tel dispositif déterministe, il y équivalence entre le fait d'être partiellement semi-local, pseudo-semi-local, quasi-semi-local et semi-local, et cela se produit si et seulement s'il existe une partition $I=J_1\cup J_2$ et deux dispositifs déterministes $g_1:\prod_{i\in J_1} Q_i\to \prod_{i\in J_1} R_i$ et $g_2:\prod_{i\in J_2} Q_i\to \prod_{i\in J_2} R_i$ tels que, à réorganisation près de l'ordre des variables dont dépendent $g_1$ et $g_2$, on ait $f=g_1\otimes g_2$.
\end{prop}

\begin{prop} Un dispositif multilocal déterministe $f:\prod_{i\in I} Q_i=Q\to R=\prod_{i\in I} R_i$ est local si et seulement si chacun des sous-dispositifs locaux $f_i$ est déterministe.
Un dispositif multilocal déterministe $f:\prod_{i\in I} Q_i=Q\to R=\prod_{i\in I} R_i$ est semi-local si et seulement s'il existe une partition $I=J_1\cup J_2$ telle que les sous-dispositifs $f_{J_1}$ et $f_{J_2}$ soient déterministes.
\end{prop}

\subsection{Structures connectives relationnelles}

Nous regroupons sous l'appellation générale de \emph{structures connectives relationnelles} toutes les structures connectives qui peuvent être associées aux dispositifs multilocaux.
Nous définissons  dans cette section  plusieurs types de structures connectives relationnelles, à savoir :

\begin{itemize}
\item diverses structures qualifiées de \emph{tensorielles},
\item deux structures dites \emph{domaniales},
\end{itemize}

et, renvoyant à un travail ultérieur, nous signalons également, sans en donner ici la définition, la notion de structure connective ludique. 

\subsubsection{Structures connectives tensorielles}\label{subsubsection structures tensorielles}

Notons respectivement par $NPS$, $NOS$, $NPL$, $NQS$, $NQL$, $NS$ et $NL$ les classes de dispositifs multilocaux d'uplicité $k\geqslant 2$ qui 
\begin{itemize}
\item ne sont pas partiellement séparables : $NPS$,
\item ne sont pas pseudo-séparables: $NOS$,
\item ne sont pas partiellement locaux: $NPL$,
\item  ne sont pas quasi-séparables: $NQS$,
\item  ne sont pas quasi-locaux: $NQL$,
\item  ne sont pas séparables : $NS$,
\item  ne sont pas locaux : $NL$.
\end{itemize}

Ainsi, nous écrirons par exemple $D\in NOS$ pour exprimer que $D$ est un dispositif multilocal d'uplicité $\geqslant 2$ et qui n'est pas pseudo-séparable.

\mbox{}

Soit maintenant $\alpha$ l'une des classes définies ci-dessus. 

\begin{df} Pour tout dispositif multilocal $D$, on appelle structure connective tensorielle de classe $\alpha$ de $D$, et on note $\kappa_\alpha(D)$, la structure connective \emph{intègre} sur l'ensemble $I=\{1,\cdots, k\}$ des indices de $D$ engendrée par les parties $J\subset I$ telles que $D_{[J]}\in\alpha$ :
\[\kappa_\alpha(D)=[\{J\in\mathcal{P}(I), D_{[J]}\in\alpha\}].\] 
\end{df}   

On pourra nommer la structure connective $\kappa_\alpha$ de la même manière que sont nommés les sous-dispositifs $D_{[J]}$ appartenant à $\alpha$. Par exemple $\kappa_{NPS}(D)$ sera appelée \emph{la structure connective non partiellement semi-locale du dispositif multilocal $D$}.

\begin{rmq} Par définition des classes considérées, l'ensemble \[\mathcal{J}=\{J\in\mathcal{P}(I), D_{[J]}\in\alpha\}\] ne contient aucune partie de $I$ de cardinal inférieur ou égal à $1$, mais la structure connective \emph{intègre} engendrée par cet ensemble, que désigne la notation $[\mathcal{J}]_1$, contient d'office la partie vide et les singletons $\{j\}$, ce qui correspond à l'intuition selon laquelle chaque lieu est connecté à lui-même. S'il a fallu traiter séparément le cas de l'uplicité $k=1$, c'est que dans ce cas particulier la localité n'est pas distincte d'une inséparable globalité.
\end{rmq}

L'implication logique indiquée à la proposition \ref{prop implications entre localites} entre les propriétés de localité de la définition \ref{df localites dispositifs} entraîne immédiatement les relations de finesse suivantes entre structures connectives  :

\begin{prop}\label{prop finesse relative des structures connectives multilocales} Pour tout dispositif multilocal $D$, on a les inclusions suivantes :
\[\kappa_{NPS}(D)\subset\kappa_{NPL}(D) \subset\kappa_{NQL}(D)\subset\kappa_{NL}(D),\]
\[\kappa_{NPS}(D)\subset\kappa_{NOS}(D) \subset\kappa_{NQS}(D)\subset\kappa_{NS}(D)\subset\kappa_{NL}(D),\] et
\[\kappa_{NQS}(D)\subset\kappa_{NQL}(D).\]
\end{prop}

\begin{exm}

\end{exm}

\subsubsection{Structures connectives domaniales}\label{subsubsection structures domaniales}

Chaque réalisation déterministe $f$ d'un dispositif multilocal $D:\prod_{i\in I} Q_i=Q\rightsquigarrow R=\prod_{i\in I} R_i$ donne lieu à une famille d'applications $(f_i:Q\to R_i)_{i\in I}$ dont chacune exprime, par son domaine de dépendance\footnote{Voir la définition \ref{df domaine indicial de dependance} page \pageref{df domaine indicial de dependance}.} $\mathrm{do}(f_i)$, une certaine connectivité entre les variables concernées vis-à-vis de $f$. Toutefois, cette connectivité peut être une sorte d'artefact propre à la fonction $f$ et non au dispositif $D$ dans son ensemble, d'où l'idée de prendre l'intersection pour toutes les réalisations déterministes de toutes les structures connectives correspondantes. C'est l'objet des deux définitions suivantes.

\begin{df}[Structure connective domaniale d'un dispositif déterministe] Soit
$f:\prod_{i\in I} Q_i=Q\to R=\prod_{i\in I} R_i$ un dispositif déterministe d'uplicité $k=card(I)$, on appelle \emph{structure connective domaniale de $f$}, et l'on note $\kappa_{do}(f)$, la structure connective intègre engendrée par les domaines de dépendance $\mathrm{do}(f_i)$ :
\[\kappa_{do}(f)=[\{\mathrm{do}(f_i),i\in I\}]_1.\]
\end{df}

\begin{xrc}
Préciser, pour un dispositif déterministe $f$, les relations de finesse entre $\kappa_{do}(f)$ et les structures de la forme $\kappa_{\alpha}(f)$ avec $\alpha\in\{NPS$, $NOS$, $NPL$, $NQS$, $NQL$, $NS$, $NL\}$.
\end{xrc}

\begin{df}[Structure connective domaniale] Soit
$D:\prod_{i\in I} Q_i=Q\to R=\prod_{i\in I} R_i$ un dispositif d'uplicité $k=card(I)$. Notant 
$\mathcal{F}(D)$ l'ensemble des réalisations déterministes de $D$, 
on appelle \emph{structure connective domaniale de $D$}, et l'on note $\kappa_{do}(D)$, l'intersection des structures connectives domaniales des applications $f\in \mathcal{F}(D)$  :
\[\kappa_{do}(D)=\bigcap_{f\in \mathcal{F}(D)}{\kappa_{do}(f)}.\] 
\end{df}

Dans les définitions précédentes, la non-localité liée au fait qu'une réponse $r_i$ serait totalement indé\-pendante de la variable $q_i$ de même indice $i$ n'est pas du tout prise en compte. Par exemple, pour le dispositif déterministe $f:\mathbf{N}^3 \to \mathbf{N}^3$ défini pour tout triplet $(q_1,q_2,q_3)\in\mathbf{N}^3$  par $f(q_1,q_2,q_3)=(q_2,q_3,q_1)$, on obtient pour $\kappa_{do}(f)$ la structure connective discrète sur $I=\{1,2,3\}$, alors même que $f$ est non locale. D'où l'idée d'introduire une autre structure connective, que nous appellerons la structure connective pointée, définie comme la structure connective domaniale à ceci près que les domaines de dépendances sont augmentés de l'indice $i$ concerné :

\begin{df}[Structure connective pointée] Soit
$f:\prod_{i\in I} Q_i=Q\to R=\prod_{i\in I} R_i$ un dispositif déterministe d'uplicité $k=card(I)$. On appelle \emph{domaine indicial de dépendance pointé} de $f$ l'ensemble d'indices noté $\mathrm{dp}(f)$ et défini par
\[\mathrm{dp}(f)=\mathrm{do}(f)\cup \{i\},\]
et on appelle \emph{structure connective pointée de $f$}, notée $\kappa_{dp}(f)$, la structure connective intègre engendrée par les domaines de dépendance pointés $\mathrm{dp}(f_i)$ :
\[\kappa_{dp}(f)=[\{\mathrm{dp}(f_i),i\in I\}]_1.\]
Plus généralement, pour tout dispositif multilocal
$D:\prod_{i\in I} Q_i=Q\to R=\prod_{i\in I} R_i$ d'uplicité $k=card(I)$, on appelle \emph{structure connective pointée de $D$}, et l'on note $\kappa_{dp}(D)$, l'intersection des structures connectives pointées des applications $f\in \mathcal{F}(D)$  :
\[\kappa_{dp}(D)=\bigcap_{f\in \mathcal{F}(D)}{\kappa_{dp}(f)},\] où comme précédemment $\mathcal{F}(D)$ désigne l'ensemble des réalisations déterministes de $D$
\end{df}

\subsubsection{Structure connective ludique}\label{subsubsection structure ludique}

Étant donnés $D:Q=Q_1\times...\times Q_k\rightsquigarrow R_1\times...\times R_k=R$ et $D':Q'=Q'_1\times...\times Q'_k\rightsquigarrow R'_1\times...\times R'_k=R'$ deux dispositifs de même uplicité $k$, une réduction de Chalons est une $2$-réduction locale telle que définie par exemple dans \cite{ChalonsRessayre:1999} ou, plus récemment, dans \cite{Chalons:201405}. Deux tels dispositifs seront dit ludiquement équivalents s'il existe entre eux une réduction de Chalons dans chaque sens. Ceci définit une relation d'équivalence sur la classe des dispositifs d'uplicité $k$, dont les classes d'équivalence sont appelées les degrés ludiques d'uplicité $k$, et la réduction de Chalons donne lieu à une relation d'ordre entre ces degrés ludiques. La classe ordonnée ainsi définie contient notamment les degrés de Tukey, qui contiennent notamment tous les ordinaux. Entre deux ordinaux quelconques, les degrés ludiques contiennent également une classe propre de degrés intermédiaires. Dans son travail, Chalons soutient que les degrés ludiques constituent un outil bien adapté à la description et à l'étude des propriétés locales et non locales de la physique quantique, en particulier de l'intrication quantique.

Or, dans un travail en cours de réalisation, Anatole Khélif associe de façon naturelle une structure connective à tout degré ludique appartenant à une classe assez large, qui contient en particulier tous les exemples qui nous intéressent ici : il y aurait donc là un procédé \og quantique\fg\, pour associer à tout dispositif multilocal, via le degré ludique auquel il appartient, une structure connective dont il faudra déterminer si elle coïncide ou non avec l'une de celle que nous avons définies précédemment.



\subsection{Ordre connectif d'un dispositif multi-local}

De manière analogue à la définition \ref{df ordre etat} de l'ordre connectif d'un état quantique, nous posons la définition suivante pour l'ordre connectif d'un dispositif multi-local.

\begin{df} Etant donné $D$ un dispositif multi-local, on appelle 
\begin{itemize}
\item \emph{ordre connectif tensoriel de $D$}   le maximum des ordres connectifs des structures tensorielles\footnote{Voir la section \ref{subsubsection structures tensorielles}.} associées à $D$,
\item \emph{ordre connectif domanial de $D$} le maximum entre l'ordre connectif de sa structure domaniale et l'ordre connectif de sa structure pointée\footnote{Voir la section \ref{subsubsection structures domaniales}.},
\item \emph{ordre connectif ludique de $D$} l'ordre connectif de sa structure connective ludique\footnote{section \ref{subsubsection structure ludique}.},
\item \emph{Ordre connectif de $D$}  le maximum entre son ordre connectif tensoriel, son ordre connectif domanial et son ordre connectif ludique.
\end{itemize}
\end{df}

\subsection{Exemples de structures connectives de dispositifs}

\begin{exm}[Dispositif $D_{EPR}$] Comme nous l'avons vu dans l'exemple \ref{exm dispositif DEPR}, ce dispositif est défini par $Q_1=Q_2=\{*\}$, $R_1=R_2=\{0,1\}$ et $D_{EPR}(**)=\{00,11\}$. Pour ce dispositif, toutes les structures connectives sont discrètes, sauf $\kappa_{NS}(D_{EPR})$ et $\kappa_{NL}(D_{EPR})$. En effet, on vérifie que $D_{EPR}$ est quasi-séparable mais non-séparable.
\end{exm}


\begin{exm}[Dispositif $D_K$] On vérifie que pour ce dispositif, décrit dans l'exemple \ref{exm dispositif DK} page \pageref{exm dispositif DK}, on a  $\kappa_{NPS}(D_K)=\kappa_{NL}(D_K)=\mathcal{B}_3$, de sorte  que toutes les structures connectives tensorielles coïncident avec la structure borroméenne. 

\begin{prop}\label{prop structure domaniale discrete de D_K}
$\kappa_{do}(D_K)$ est la structure connective discrète.
\end{prop}
\textbf{Preuve}. Cela découle de l'existence de réalisations déterministes de $D_K$ qui ne dépendent que de deux quelconques des trois variables en jeu. C'est par exemple le cas de l'application définie de la façon suivante :
\[
f(00q_3)=000, f(01q_3)=011, f(10q_3)=101, \,\mathrm{et}\, f(11q_3)=111,
\]
pour laquelle $\mathrm{do}(f_1)=\mathrm{do}(f_2)=\mathrm{do}(f_3)=\{1,2\}$. 
\begin{flushright}
\textbf{CQFD}
\end{flushright}

\begin{prop}
$\kappa_{dp}(D_K)=\mathcal{B}_3$.
\end{prop}
\noindent\textbf{Preuve}.
La fonction $f$ considérée dans la preuve de la proposition \ref{prop structure domaniale discrete de D_K}
montre que $\{1,3\}$ et $\{2,3\}$ n'appartiennent pas à $\kappa_{dp}(GHZ)$. On vérifie de même que $\{2,3\}$ n'y est pas non plus.
On vérifie par l'absurde que $I\in \kappa_{dp}(D_K)$ car sinon il existerait néces\-saire\-ment une réalisation déterministe $f$ de $D_K$ pour laquelle il existerait un singleton dans $I$ tel que chacun des domaines pointés des $f_i$ soient soit inclus soit dans ce singleton, soit inclus dans son complémentaire. Si, par exemple, un tel singleton pouvait être pris égal à  $\{3\}$, on devrait nécessairement avoir avoir $do(f_1)\subset \{1,2\}$, $do(f_2)\subset \{1,2\}$ et $do(f_3)\subset \{3\}$. Nous laissons au lecteur le soin de vérifier qu'aucune réalisation déterministe du dispositif $D_K$ ne peut en fait vérifier cette condition.
\begin{flushright}
\textbf{CQFD}
\end{flushright}
\end{exm}

On trouvera d'autres exemples de structures connectives de dispositifs multilocaux dans la présentation placée sur la page web

\url{https://sites.google.com/site/logiquecategorique/Contenus/CSQE}.

\section{Structures connectives des dispositifs probabilistes} \label{section dispositifs probabilistes}

Dans les considérations précédentes, nous avons laissé de coté les aspects proprement probabilistes des expériences de mesure d'un système quantique, puisque ce que nous avons appelé un dispositif multilocal est défini par l'ensemble des réponses \emph{possibles} du système, indépendamment de leur probabilité.


Grâce à la notion de structure connective d'une famille finie de variables aléatoires, dont nous devons l'idée à Anatole Khélif, tout \emph{dispositif multi-local probabiliste} associé à une expérience de mesure sur un système intriqué donne également lieu à une structure connective. Dans le présent article, nous ne définirons pas formellement les dispositifs multi-locaux probabilistes, nous contentant de donner la définition de la structure connective d'une famille finie de variables aléatoires. L'idée consiste à définir, pour toute famille finie $(X_i)_{i\in I}$ de variables aléatoires définies sur un même espace probabilisé $(\Omega, P)$ et à valeurs dans des espaces mesurables $R_i$ :
\[X_i:\Omega \rightarrow R_i,\] le fait d'être séparable :

\begin{df}  $(X_i)_{i\in I}$ est \emph{séparable} s'il existe une partition $I=I_1\cup I_2$ en deux parties disjointes non vides, telle que les deux variables aléatoires  $Y_n:\Omega\rightarrow \prod_{i\in I_n} R_i$, avec $n\in\{1,2\}$, soient indé\-pendantes, où chaque $Y_n$ est définie par
\[Y_n=(X_i)_{i\in I_n}.\]

\end{df}

Autrement dit, pour toute partie mesurable $A_1\subset \prod_{i\in I_1} R_i$ et toute partie mesurable $A_2\subset \prod_{i\in I_2} R_i$, on doit avoir 
\[P((X_i)_{i\in I}\in A_1\times A_2)= P((X_i)_{i\in I_1}\in A_1)P((X_i)_{i\in I_2}\in A_2).\]


\begin{df} La \emph{structure connective de la famille de variables aléatoires $(X_i)_{i\in I}$}  est la structure connective sur $I$ engendrée par les parties $J\subset I$ pour lesquelles $(X_j)_{j\in J}$ est une famille non séparable de variables aléatoires.
\end{df}

Pour une expérience de mesure sur un système intriqué, il s'agit alors d'appliquer la définition ci-dessus à une certaine famille de variables aléatoires associée au dispositif quantique concerné.

\begin{xrc} L'ensemble des parties $J$ pour lesquelles  $(X_j)_{j\in J}$ est une famille non séparable de variables aléatoires forme-t-il une structure connective ? Autrement dit, dans la définition précédente, aurait-on  pu remplacer le mot \og engendrée\fg\, par \og constituée\fg ?
\end{xrc}

\begin{exm} [Trois variables aléatoires borroméennement connectées]\label{exm trois variables aleatoires borro} Étant donné un entier $n\geqslant 2$, étant données deux variables aléatoires $X_1$ et $X_2$ indé\-pendantes prenant avec des probabilités uniformes leurs valeurs dans $\{0,1,\cdots, n-1\}$ , appelons $X_3$ la variable aléatoire obtenue en faisant la somme dans $\mathbf{Z}/n\mathbf{Z}$ de $X_1$ et $X_2$. On vérifie facilement que ces trois variables aléatoires sont deux à deux séparables, mais que la famille qu'elles forment toutes trois ne l'est pas. Par conséquent, la structure connective de $(X_1,X_2,X_3)$ est borroméenne. Par exemple, pour $n=2$, on peut prendre $\Omega=\{(0,0), (0,1),(1,0), (1,1)\}$ muni de la probabilité uniforme, $X_1$ telle que $X_1(a,b)=a$, $X_2$ tel que $X_2(a,b)=b$, et $X_3(a,b)\equiv a+b \,[\mod 2\,]$.
\end{exm}

\begin{exm} [Familles de variables aléatoires à structure connective brunnienne]\label{exm n variables aleatoires brunniennes}
Plus généralement, pour tout entier $k\geqslant 1$, il suffit de prendre $k$ variables aléatoires $X_1$, $\cdots$, $X_k$ indé\-pendantes et identiquement distribuées selon la loi uniforme sur $\mathbf{Z}/n\mathbf{Z}$ et de poser $X_{k+1}\equiv\sum_{i=1}^{i=k} X_i \,[\mod n\,]$ pour obtenir une famille $(X_1,\cdots,X_{k+1})$ de structure brunnienne : ces variables sont globalement connectées, mais à part les singletons toute sous-famille non vide est séparable.
\end{exm}

En formant des vecteurs aléatoires avec ces structures brunniennes, on vérifie facilement qu'on peut construire des familles de variables aléatoires admettant n'importe quelle structure connective donnée :

\begin{prop}
Toute structure connective finie est celle d'une famille de variables aléatoires.
\end{prop}

\section{Conclusion et remerciements}

Une grande diversité de structures connectives a été associée dans le présent travail aux états quantiques intriqués et aux dispositifs de mesure portant sur ces états, permettant en particulier de leur associer un ordinal, l'ordre connectif, constituant ainsi autant d'outils de description et de classification de l'intrication quantique. L'étude approfondie de ces structures, et en particulier des relations entre elles, reste à mener. Il reste également à prouver les diverses \emph{conjectures brunniennes}\footnote{Voir la section \ref{subsection conjectures brunniennes} page \pageref{subsection conjectures brunniennes}.} associées à ces structures. 

\textbf{Remerciements}. Je remercie les participants du séminaire de logique caté\-gorique de Paris 7, en particulier Anatole Khélif, Saab Abou-Jaoudé, Christophe Chalons ainsi que Jean-Jacques Rozenbaum, Marc Lachièze-Rey et Jean-Jacques Szczeciniarz, ceux du \emph{Workshop on Diffeology, etc.} qui s'est tenu à Aix-en-Provence en juin 2014, en particulier Patrick Iglesias-Zemmour et Pierre Cartier, et les organisateurs du séminaire Quartz-LISMMA (Supméca Paris, juillet 2014) pour les échanges d'idées et les encouragements reçus. 

\section{Index des notations}

\begin{itemize}
\item $[\mathcal{A}]$ ou $[ \mathcal{A}]_1$, structure connective intègre engendrée par un ensemble $\mathcal{A}$ de parties d'un ensemble de points : section \ref{notation structure connective engendree} page \pageref{notation structure connective engendree},
\item $b_J$, projection sur $J\subset I$ d'un état $b$ $(J,\neg J)$-séparable, section \ref{notation bJ} page \pageref{notation bJ},
\item $\exists^\bullet$, quantificateur \og certains mais pas tous\fg, formule (\ref{formule quantificateur pas tous}) page \pageref{formule quantificateur pas tous},
\item  $\neg \mathcal{D}_J$, ensemble des $J$-états intriqués, formule (\ref{notation neg mathcal DJ}) page \pageref{notation neg mathcal DJ},
\item $\mathcal{D}_{J}$, ensemble des $J$-états séparables, définition \ref{df J-etats separables} page \pageref{df J-etats separables},
\item $\mathcal{D}_{(J_1,J_2)}$, ensemble des $J$-états $(J_1,J_2)$-séparables : définition \ref{notation etats J1J2 separables} page \pageref{notation etats J1J2 separables},
\item $H_J$, espace des $J$-états, formule (\ref{notation HJ}) page \pageref{notation HJ},
\item $\neg J$, complémentaire de $J$ dans $I$, section \ref{notation negJ} page \pageref{notation negJ},
\item $\mathcal{L}(H)$, espace des endomorphismes de $H$,
\item $M[a]$, ensemble des états accessibles à partir de $a$ après une mesure $M$, formule (\ref{formule notation ensemble des etats apres mesure}) page \pageref{formule notation ensemble des etats apres mesure},
\item $M_J[a]$ ensemble des projections sur $J$ des états accessibles à partir de $a$ après une expérience $M$ déterminante sur $\neg J$ , formule (\ref{notation MJa}) page \pageref{notation MJa},
\item  $\mathcal{M}_L$, ensemble des expériences déterminantes sur $L\subset I$, section \ref{notation mathcalM L} page \pageref{notation mathcalM L},
\item $R:A\rightsquigarrow B$, relation binaire de $A$ vers $B$ vue comme application multivalente, formule (\ref{formule relation binaire comme transition}) page \pageref{formule relation binaire comme transition},
\item $\widetilde{\mathcal{S}}_\mathcal{F}$, ensemble des états du système $\mathcal{S}$ (dans le formalisme $\mathcal{F}$), définition \ref{df systeme quantique} page \pageref{df systeme quantique},
\item $\mathcal{T}(H)$, ensemble des opérateurs positifs sur $H$ et de trace $1$, définition \ref{df notation TH} page \pageref{df notation TH},
\item $<v\vert A\vert w>$ : remarque \ref{rmq sur ecriture braket v A w} page \pageref{rmq sur ecriture braket v A w},
\item $\Omega(\kappa)$, ordre connectif de la structure connective intègre $\kappa$ : section \ref{subsection ordre connectif}, page \pageref{subsection ordre connectif},
\item $\Omega(\psi)$, ordre connectif de l'état quantique (pur) $\psi$ : formule (\ref{formule ordre connectif etat pur}) page \pageref{formule ordre connectif etat pur}.
\end{itemize}

\newpage


\bibliographystyle{plain}


\newpage

\tableofcontents

\end{document}